**Dispersion interaction of two graphene sheets**

M.V. Davidovich

Saratov N.G. Chernyshevsky National Research State University

E-mail: DavidovichMV@info.sgu.ru

The Casimir method for determining the dispersive force by varying zero vacuum energy fluctuations is applied to two graphene sheets in the approximation of the Drude model for surface conductivity. As an alternative, the Van Kampen summation method is used. The force is determined for small and for large distances between the sheets. The results of both models are quite similar. Precisely, at large distances, the attractive force decreases inversely proportional to the fourth power of the distance. At short distances, the force is a finite attractive one. With a small chemical potential, the force can have a minimum at distances of the order of 0.3 nm, then increases, reaches a maximum at distances of the order of 200 nm, and at large distances decreases inversely proportional to the fourth power of the distance. At a chemical potential of significantly more than 1 eV, a minimum is not observed.

Two parallel graphene sheets at a distance $d$ between them and at zero temperature can interact with the Casimir force, determined similarly to [1], namely by summing the natural frequencies of a resonator with ideal walls Fig. 1 and with an electric or magnetic wall in the center:

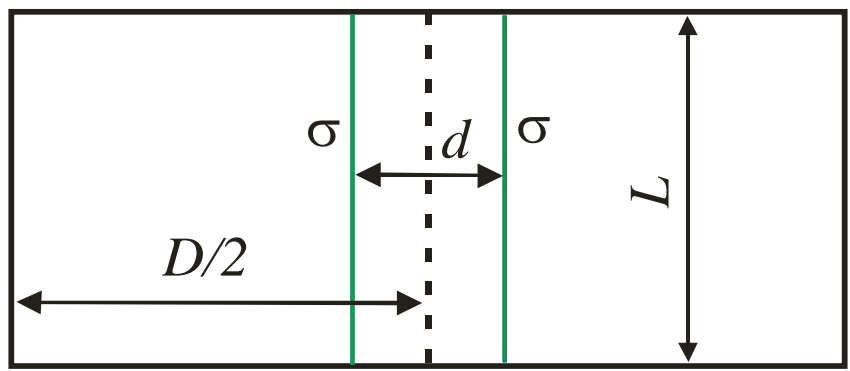

Fig. 1. Rectangular resonator with two conductive graphene sheets. The dash marks the walls: electric or magnetic

$$\tilde{E}(d) = \frac{\hbar}{2} \text{Re} \sum_{\substack{\alpha=e,h \\ \beta=e,h}} \sum_{mnl} \tilde{\omega}_{mnl}^{\alpha\beta}(d). \qquad (1)$$

Here, the tilde indicates the frequencies perturbed by graphene. In an empty resonator, there are undisturbed resonant frequencies $\omega_{mnl}^h = c\sqrt{k_{xm}^2 + k_{yn}^2 + k_{zl}^2}$ of TE$_{mnl}$ (or H$_{mnl}$) modes, where $k_{xm} = m\pi/L$, $k_{yn} = n\pi/L$, $k_{zl} = l\pi/D$, $m=0,1,...$, $n=0,1,...$, $l=1,2,...$, except for $m=n=0$, as well as frequencies of TM$_{mnl}$ (or E$_{mnl}$) modes, the difference of which is that now that $m=1,2,...$, $n=1,2,...$, $l=0,1,2,...$[2,3]. Thus, oscillation degeneracy takes place in an empty resonator. It is removed in a resonator filled with graphene. Going to the limit $L \to \infty$ means continuity of the transverse indices $dk_{xm} = dk_x = (\pi/L)dm$, $dk_{yn} = dk_y = (\pi/L)dn$, and replacing the two-dimensional sum in (1) with a two-dimensional integral. Next, it is convenient to switch to the polar coordinates $k_x = \kappa\cos(\varphi)$, $k_y = \kappa\sin(\varphi)$. The angle integral is calculated and is equal to $2\pi$. The remaining sum over $l$ can be calculated, but it is easier to go to the limit $D \to \infty$, $dk_{zl} = dk_z = (\pi/D)dl$ and reduce it to an integral. Then the sum (1) reduces to a two-dimensional integral when the resonator is expanded to the entire space. With a limiting transition to the entire space in (1), we have equality $k_0 = \sqrt{\kappa^2 + k_z^2}$. At the same time, always $k_z^2 \leq k_0^2$. This means that only the radiated modes are taken into account. They are the only ones in the final resonator. In the infinite space along the graphene sheets, slow surface plasmons are possible. For them $k_z^2 < 0$ (in the absence of dissipation), i.e. these modes decay exponentially in the longitudinal direction. They should be taken into account when interacting graphene sheets. They are not present in the Casimir problem, since plasmons are not possible along a perfectly conducting surface. The derivative $\partial_d \tilde{E}(d)$ means force, and if it is positive, then the force is attractive, since energy increases with distance. For the Casimir pressure considered between the sheets, one should take $P(d) = -\partial_d \tilde{E}(d)/L^2$. In its meaning, it determines the pressure between the layers. Negative pressure means attraction. The characteristic equation for two sheets in a finite-size resonator has the form

$$y_{mn}^{e,h} + i\left(y_{mn}^{e,h}\left[i\tan(k_z d/2)\right]^{-s} + \varsigma\right)\tan(k_z(D-d)/2) = 0. \qquad (2)$$

It contains the normalized (dimensionless) conductivity of graphene $\varsigma = \sigma\eta_0$, $\eta_0 = \sqrt{\mu_0/\varepsilon_0} = 376.78$ (Ohms) are the characteristic vacuum resistance, $y_{mn}^e = k_0/k_z = k_0/\sqrt{k_0^2 - k_{xm}^2 - k_{yn}^2}$, $y_{mn}^h = k_z/k_0 = \sqrt{k_0^2 - k_{xm}^2 - k_{yn}^2}/k_0$ are the normalized conductivity modes, $s = (-1)^\nu$, $\nu=1$ ($s=-1$) corresponds to the magnetic wall, $\nu=2$ ($s=1$) corresponds to the electric wall. All modes are divided into modes with electric and magnetic walls in the center of Fig. 1. In this case, the resonance condition in the empty resonator for the electric wall $k_z = k_{zle} = 2l\pi/D$ and for the magnetic wall $k_z = k_{zlh} = (2l-1)\pi/D$ leads to the condition $k_0 = \omega_{mnl}^{\alpha,\beta}/c = \sqrt{k_{xm}^2 + k_{yn}^2 + k_{zl\beta}^2}$. Here $\beta = e$ means an electric wall, $\beta = h$ – a magnetic wall, and – $\alpha = e, h$ an electric or magnetic mode. Obviously, this is the case when $\varsigma = 0$. In the other extreme case $\varsigma \to \infty$, it should be $\tan(k_z(D-d)/2) = 0$ either $k_z = 2l\pi/(D-d)$ for both walls. These are the frequencies of the two outer parts of the resonator. For the inside $k_z = l\pi/d$. Obviously, the force does not act on such ideal screens if $d = D/3$. The Casimir pressure in this case is always infinite due to the divergence of the series

$$P(d) = -\partial_d \delta E = -\frac{\hbar c \pi^2}{d^3} \sum_{m,n,l} l^2 / \sqrt{\left(\frac{m\pi}{L}\right)^2 + \left(\frac{n\pi}{L}\right)^2 + \left(\frac{l\pi}{d}\right)^2}.$$

However, when $d \to \infty$ we get the ratio of two infinite quantities, from which we can isolate the final result. In the presence of graphene, the characteristic equations are written as

$$\tan\left(\tilde{k}_{zle}\frac{D-d}{2}\right) = \alpha_{e,h}(\tilde{k}_{zle}) = -\frac{y_{mn}^{e,h}\tan(\tilde{k}_{zle}d/2)}{y_{mn}^{e,h} + i\varsigma\tan(\tilde{k}_{zle}d/2)}, \tag{3}$$

$$\tan\left(\tilde{k}_{zlh}\frac{D-d}{2}\right) = \beta_{e,h}(\tilde{k}_{zlh}) = -\frac{y_{mn}^{e,h}}{i\varsigma - y_{mn}^{e,h}\tan(\tilde{k}_{zlh}d/2)}. \tag{4}$$

In them, the tilde indicates the values perturbed by graphene, and $\varsigma$ the normalized conductivity of graphene. It is easy to see that when $\varsigma = 0$ we obtain the modes of an empty resonator. We will look for solutions to these equations in the form $\tilde{k}_{zle} = k_{zle} + \Delta\tilde{k}_{zle}^{e,h}$ and $\tilde{k}_{zlh} = k_{zlh} + \Delta\tilde{k}_{zlh}^{e,h}$, where $k_{zle} = 2l\pi/D$, $k_{zlh} = (2l-1)\pi/D$. It follows from equation (3) that there is a solution $\Delta\tilde{k}_{z0e}^{e,h} = 0$ for $l = 0$. This mode does not contribute to the force, therefore, we further consider $l = 1, 2, \ldots$ Equations (3), (4) to take the form

$$\tilde{k}_{zle}^{(e,h)} = \frac{2l\pi}{D} + \Delta\tilde{k}_{zle}^{(e,h)} = \frac{2l\pi}{D-d} + \frac{2\arctan(\alpha_{e,h}(\tilde{k}_{zle}^{(e,h)}, d))}{D-d}. \tag{5}$$

$$\tilde{k}_{zlh}^{(e,h)} = \frac{(2l-1)\pi}{D} + \Delta\tilde{k}_{zlh}^{(e,h)} = \frac{(2l-1)\pi}{D-d} + \frac{2\arctan(\beta_{e,h}(\tilde{k}_{zlh}^{(e,h)}, d)) + \pi}{D-d}. \tag{6}$$

In these equations, at the limiting transition , it is possible to replace $D - d \to D$. These are implicit equations, and their solution can be obtained by iteration, taking the initial values $k_{zle}$ and $k_{zlh}$. With this solution, we obtain power expansions of a small parameter $1/D^n$, which disappear at $D \to \infty$ except $n = 1$. Indeed, the double iteration (5) gives

$$\tilde{k}_{zle}^{(e,h)} = k_{zle}^{(e,h)} - \frac{2\arctan\left(\frac{y_{mn}^{e,h} \tan(k_{zle}^{(e,h)}d/2)[1+\delta_1 d/D]}{\left(y_{mn}^{e,h} + i\varsigma \tan(k_{zle}^{(e,h)}d/2)\right)[1+\delta_2 d/D]}\right)}{D},$$

$$\delta_1 = \frac{y_{mn}^{e,h} \arctan\left(\alpha_{e,h}\left(k_{zle}^{(e,h)},d\right)\right)}{\cos^2(kd/2)y_{mn}^{e,h} \tan(kd/2)},$$

$$\delta_2 = \frac{i\varsigma \arctan\left(\alpha_{e,h}\left(k_{zle}^{(e,h)},d\right)\right)}{D\cos^2\left(k_{zle}^{(e,h)}d/2\right)\left(y_{mn}^{e,h} + i\varsigma \tan(k_{zle}^{(e,h)}d/2)\right)}.$$

At the limit transition $D \to \infty$, the indices become continuous variables $k_{zle}^{(e,h)} \to k_z = k$, square brackets turn into one, and we get $\tilde{k} = k + 2\arctan(\alpha(k,d))/D$, moreover, $k$ is a single variable, and equations (5) and (6) coincide when replaced $\alpha \leftrightarrow \beta$. It is convenient to denote in (3) and (4) $i\varsigma = 1/(k_0\delta)$ where the complex parameter $\delta$ has the dimension of length. Since we next move from summation over l to integration over $dl = Ddk_z/(2\pi)$, all the terms of the expansion except the first one disappear $D \to \infty$, and we get

$$\Delta\tilde{k}_{ze}^{e,h}(d,\delta) = \frac{2\arctan(\alpha_{e,h}(k,d,\delta))}{D}, \tag{7}$$

$$\Delta\tilde{k}_{zh}^{e,h} = \frac{2\arctan(\beta_{e,h}(k,d,\delta))}{D}, \tag{8}$$

$$\alpha_{e,h}(k,d,\delta) = -\delta \frac{k_0 y_{mn}^{e,h} \tan(kd/2)}{\delta k_0 y_{mn}^{e,h} + \tan(kd/2)}, \tag{9}$$

$$\beta_{e,h}(k,d,\delta) = -\frac{\delta k_0 y_{mn}^{e,h}}{1 - \delta k_0 y_{mn}^{e,h} \tan(k_{zlh}d/2)}. \tag{10}$$

For discrete quantities, we introduce the notation $\tilde{k}_0 = \tilde{k}_{mnl} = \sqrt{k_{xm}^2 + k_{yn}^2 + \tilde{k}_{zl(e,h)}^2}$, $k_{xm}^2 + k_{yn}^2 = \kappa^2$, $\Delta_{\alpha\beta} = \left(\Delta\tilde{k}_{zl\beta}^\alpha\right)^2 + 2k_{zl\beta}\Delta\tilde{k}_{zl\beta}^\alpha$, $\tilde{k}_0 = k_0\sqrt{1 + \Delta_{\alpha\beta}/k_0^2}$, and for the perturbed frequencies we get

$$\tilde{\omega}_{mnl}^{\alpha\beta}(d) = \omega_{mnl}^{\alpha\beta}\sqrt{1 + \Delta_{(\alpha\beta)}(d)/k_0^2}, \tag{11}$$

and for the difference between perturbed and undisturbed frequencies, we have

$$\tilde{\omega}_{mnl}^{\alpha\beta}(d) - \omega_{mnl}^{\alpha\beta} = \omega_{mnl}^{\alpha\beta}\left(\sqrt{1 + \Delta_{\alpha\beta}/k_{mnl}^2} - 1\right).$$

Summing frequencies (11) gives infinite energy (1), as well as summing perturbations. The attractive force acting on a graphene sheet is obtained by differentiation

$$F(d) = \frac{\hbar}{2}\text{Re}\frac{\partial}{\partial d}\sum_{\substack{\alpha=e,h \\ \beta=e,h}}\sum_{mnl}\left(\tilde{\omega}_{mnl}^{\alpha\beta}(d) - \omega_{mnl}^{\alpha\beta}\right) = \frac{\hbar}{2}\text{Re}\sum_{\substack{\alpha=e,h \\ \beta=e,h}}\sum_{mnl}\frac{\partial}{\partial d}\tilde{\omega}_{mnl}^{\alpha\beta}(d), \tag{12}$$

and the pressure (force per unit area) is

$$P(d) = \frac{F(d)}{L^2} = \frac{\hbar}{2L^2} \text{Re} \sum_{\substack{\alpha=e,h \\ \beta=e,h}} \sum_{mnl} \frac{\partial}{\partial d} \tilde{\omega}_{mnl}^{\alpha\beta}(d). \qquad (13)$$

This result has the form of a ratio of two infinite quantities, and the final result can be extracted from it. Aiming as in [1] $L \to \infty$, let's move from summation by transverse indices to integration. In this case $dk_{xm} = (\pi/L)dm$, $dk_{yn} = (\pi/L)dn$. With an extremely large size $L$, the indexes run through continuous values. For integration, we introduce polar coordinates $\kappa^2 = k_{xm}^2 + k_{yn}^2$, $dmdn = (L/\pi)^2 dk_{xm} dk_{yn}$, $dk_{xm} dk_{yn} = d\kappa d\varphi$. Summation by the longitudinal index is also replaced by integration $dl = (D/2\pi)dk_{zl}$. For any finite $D$, you can calculate the sum and determine the strength. Given the value of the angle integral $2\pi$, we have the final result

$$P^{\alpha\beta}(d,\delta) = \frac{F^{\alpha\beta}(d,\delta)}{L^2} = \frac{\hbar c}{2\pi^2} \text{Re} \int_0^\infty \int_0^\infty \Phi^{\alpha\beta}(k,d,\delta) \frac{\kappa d\kappa k dk}{\sqrt{\kappa^2 + k^2}}, \qquad (14)$$

$$\Phi^{\alpha e}(k,d,\delta) = 2\frac{\partial_d \alpha_\alpha(k,d,\delta)}{1+\alpha_\alpha^2(k,d,\delta)} + k,$$

$$\Phi^{\alpha h}(k,d,\delta) = 2\frac{\partial_d \beta_\alpha(k,d,\delta)}{1+\beta_\alpha^2(k,d,\delta)} + k.$$

By replacing variables $k = \sqrt{k_0^2 - \kappa^2}$, $kdk = k_0 dk_0$, we transform (14) to the form

$$P^{\alpha\beta}(d,\delta) = \frac{\hbar c}{2\pi^2} \text{Re} \int_0^\infty \int_0^\infty \Phi^{\alpha\beta}(k_0,\kappa,d,\delta) \kappa d\kappa dk_0. \qquad (15)$$

If we do not separate the modes with the electric and magnetic walls, but use the equations for the entire resonator, then we get the equations $\tan(\tilde{k}_{zl} D/2) = \alpha_{e,h}(\tilde{k}_{zl})$ in which

$$\alpha_e = -\frac{(k_0^2 \delta)^2 \tan(kD/2) + k_0^2 \delta k \tan(kd)[k_0^2 \delta + k \tan(kD/2)]}{(k_0^2 \delta)^2 + 2k_0^2 \delta k \tan(k_z D/2) + \tan(kd)[(k^2 - (k_0^2 \delta)^2)\tan(k_z D/2) + k_0^2 \delta k]},$$

$$\alpha_h = -\frac{(\delta k)^2 \tan(kD/2) + k\delta \tan(kd)[k\delta + \tan(kD/2)]}{(k\delta)^2 + 2k\delta \tan(kD/2) + \tan(kd)[(1-(k\delta)^2)\tan(kD/2) + k\delta]}.$$

Relations (14), (15), obtained similarly to the summation in [1], do not take into account the contribution of evanescent (attenuating) inhomogeneous plane waves, since the value $k_z$ is always real (ignoring dissipation). There are no such waves in the final resonator. They are also absent in the Fabry-Perrault resonator made of perfectly conductive screens, which corresponds to the Casimir problem. However, in a free space with two graphene sheets, they must be taken into account. These waves are strongly attenuated in the z direction, so formulas (14) and (15) should give correct results at large distances. To account for the evanescent contributions, we replace $k = \sqrt{k_0^2 - \kappa^2}$, $kdk = k_0 dk_0$ and transform the integrals (14), (15) into

$$\int_0^\infty dk_0 \left( \int_0^{k_0} \Phi^{\alpha\beta}(k_0,\kappa,d,\delta)\kappa d\kappa + \int_{k_0}^\infty \Phi^{\alpha\beta}(k_0,\kappa,d,\delta)\kappa d\kappa \right), \qquad (16)$$

where is in the first integral $k = \sqrt{k_0^2 - \kappa^2}$, and in the second integral $k = -i\sqrt{\kappa^2 - k_0^2} = -iK$. Obviously, the first integral in (16) coincides with (15), and the second gives an addition. Then the modified formulas allow us to calculate the force at short distances using the graphene model as a two-dimensional conductive sheet. Unfortunately, measuring force at short distances of the order of nanometers is still very problematic, so such models are of interest. Although the dispersion force for dielectrics is mainly determined by their low-frequency properties [4,5], its strict definition requires knowledge of the dispersion properties over the entire range. The use of simple analytical models of dispersion of the Drude-Lorentz type does not allow us to accurately describe the dispersion. In particular, it significantly depends on the internal field. Although for some substances, for example, diamond, the dielectric constant can be determined from experimental data on losses [6] in a wide range, the use of such numerical results to determine the force in complex integral formulas is very problematic. To determine the force over long distances, it is sufficient to know the dispersion at low frequencies [4]. For graphene, it is possible to introduce a conduction model that describes its properties well enough in a wide frequency range. For further study of integrals, it is necessary to introduce such a model. A significant number of papers have been devoted to graphene conductivity models, for example, [7–19]. Graphene is a highly nonlinear material, but for weak dispersion fluctuations, its linear response is important. Most linear response models use approximate approaches to determining conductivity, in particular, the Dirac approximation of massless fermions, leading to scalar conductivity. This type of model is also used in work on dispersion forces in graphene [20–27]. The strict linear conductivity of graphene is tensor, and taking into account spatial dispersion is given by formula (8) of [13]. It contains intraband and interband conduction. All such models include the collision frequency (CF) $\omega_c$ (or relaxation time $\tau_r = 1/\omega_c$) – a phenomenological parameter. Its introduction significantly affects the conductivity. At low frequencies, it is usually considered constant, but obviously it is a function of frequency. Graphene's anisotropy is weak, and spatial dispersion is manifested at large wave vectors (frequencies), when the contribution to the dispersive force is already small. Therefore, we will use a model of intraband scalar conductivity of the Drude type throughout the entire range

$$\sigma(\omega) = \frac{\sigma(0)}{1 + i\omega/\omega_c(\omega)},$$

considering the CF as a function of frequency. The approximate integration of the approximate formula of linear dispersion in infinite Dirac cones for intraband conductivity leads to the form

$$(16) \quad \sigma_{intra}(\omega, \omega_c, \mu_c, T) = \frac{ie^2}{\pi\hbar^2(\omega - i\omega_c)} \int_0^\infty (\partial_\varepsilon f_{FD}(\varepsilon) - \partial_\varepsilon f_{FD}(-\varepsilon))\varepsilon d\varepsilon,$$

namely [16]

$$\sigma_{intra}(0, T) = \frac{4\sigma_0 k_B T}{\pi\hbar\omega_{c0}} \ln\left(2 + 2\cosh\left(\frac{\mu_c}{k_B T}\right)\right), \qquad (17)$$

where $\sigma_0 = e^2/(4\hbar) = 6.085 \cdot 10^{-5}$ S, $\sigma(0) = 4\sigma_0\mu_c/(\pi\hbar\omega_{c0})$, $f_{FD}(\varepsilon)$ is the Fermi–Dirac function for temperature $T$ and chemical potential $\mu_c$, $\omega_{c0}$ is the low–frequency CF. Assuming $\omega_{c0} = 0$ that we obtain at zero temperature $\sigma(0) = 4\sigma_0\mu_c/(\pi\hbar)$, $\sigma(\omega) = -i\sigma(0)/\omega$, which leads to actual resonant frequencies in (1) and to the absence of the need to take the real part. In pure graphene, the CF rate is very low at zero temperature. As shown above, frequencies $\omega > \omega_{c0}$ make the main contribution to the force. Therefore, it is quite possible to omit the small imaginary parts in (1). For interband conductivity in this approximation, we have

$$\sigma_{inter}(\omega, \mu_c, \omega_{c0}) = \frac{-4i\sigma_0(\omega - i\omega_{c0})}{\pi\hbar} \int_0^\infty \frac{f_{FD}(-\varepsilon) - f_{FD}(\varepsilon)}{(\omega - i\omega_{c0})^2 - (2\varepsilon/\hbar)^2} d\varepsilon.$$

The contribution of this part of the conductivity at low frequencies is very small and significant at optical frequencies, which almost do not contribute to the force. At high frequencies, it decreases as $1/\omega$, i.e. it is also described by a formula like (16). We'll ignore it by counting $\sigma(0) = \sigma_{intra}(0)$. In the field of an electromagnetic wave with an amplitude $E_0$ the almost free $\pi$-electron of graphene with an effective mass $m_{eff}$ oscillates at a distance $r = (eE_0/m_{eff})\cos(\omega t)/\omega^2$ [28]. It determines the CF. At frequency $\omega > \omega_p = \gamma_0/\hbar$, the photoionization occurs, bonds break, $m_{eff} = m_e$, and with weak electromagnetic fluctuations, the electron displacement may be less than the lattice constant: $r < a = 0.246$ nm. Here, $\gamma_0 = 3$ eV is the coupling constant. At high frequencies, the offset may become significantly less than the low-frequency free path. This means that the frequency of collisions decreases with frequency at least according to the law $1/\omega^2$. The characteristic frequency near which a strong change of CF occurs is $\omega_p = 4.57 \cdot 10^{15}$ Hz (the corresponding wavelength is $\lambda_p$=458 nm). On it, the $\pi$-bonds break, $m_{eff} = m_e$, and the electron oscillates around the carbon atom, experiencing almost no scattering. In the X – ray range it can be considered $\omega_c = 0$. At low frequencies the CF does not depend on the frequency. Therefore, at zero temperature we have a model

$$\omega_c(\omega) = \frac{\omega_{c0}\omega_p^\nu}{\omega_p^\nu + \omega^\nu}.$$

$$\sigma(\omega) = \frac{\sigma(0)\omega_{c0}\omega_p^\nu}{\omega_{c0}\omega_p^\nu + i\omega(\omega_p^\nu + \omega^\nu)}. \tag{18}$$

$$\sigma(0) = 4\sigma_0\mu_c / (\pi\hbar\omega_{c0}).$$

We will use $\nu \geq 3$ what ensures convergence. Calculations were performed for $\nu=4$. In this model, the force depends on the magnitude $\mu_c / (\hbar\omega_{c0})$. If $T \to 0$, the CF and the chemical potential in pure graphene tend to zero, but in such a way that their ratio is finite, then the force does not disappear. Instead of the Drude model, it is often assumed $\omega_{c0} = 0$ in metal films to avoid dissipation [29–31]. This model is called a plasma model. In this case, there is also no dissipation for graphene: $\sigma(\omega) = -4i\sigma_0\mu_c / (\pi\hbar\omega)$, and all frequencies are real. We will use a small CF, taking the real parts from relations of type (11). This issue has been discussed quite extensively in a number of papers (see [29–31]). In particular, in [29] it is proposed to omit imaginary parts. In [32], a method was developed for generalizing Casimir forces to the case of interaction of a dissipative quantum oscillator with a thermostat in the framework of the Zwanzig–Caldeira–Leggett theory. However, in our case, there is no thermal field, and the classical approach can be used [1]. In thermodynamic equilibrium, including the region near zero temperature, dissipation means that at each frequency the body absorbs exactly as much as it emits, therefore at each frequency the total energy of the resonator is equal to the real part of the complex energy, as in the case of the monochromatic process [2,3]. Next, the Van Kampen method with an imaginary frequency will be used, which does not lead to complex quantities. For the model (16) at ultra-low frequencies lower $\omega_{c0}$, the conductivity is real. This area contributes almost nothing to the force. At low frequencies, which are large then $\omega_{c0}$ there is the main contribution to the force, the conductivity (3) is imaginary inductive $\sigma(\omega) = -i\sigma(0)\omega_{c0}/\omega$ and decreases with frequency. At very high frequencies $\delta(\omega,T) \to \infty$, where the conductivity also makes almost no contribution to the force, it is imaginary and decreases as $1/\omega^{\nu+1}$. Temperature accounting is reduced to replacing $\sigma(0) \to \sigma(0,T)$ where $\sigma(0,T) = \sigma_{\text{intra}}(0,T)$ (17) should be used [16,19]. Next, we use the dimensionless conductivity of graphene $\varsigma(\omega,T) = \sqrt{\mu_0/\varepsilon_0}\sigma(\omega,T)$. We introduce a complex quantity $\delta(\omega,T) = 1/[ik_0\varsigma(\omega,T)]$ with the dimension of length. At low frequencies it does not depend on the frequency (for low dissipation). At high frequencies $\delta(\omega,T)\to\infty$. Turning to the wave numbers $k_{c0}=\omega_{c0}/c$, $k_p=\omega_p/c$, $k_0=\omega/c$, the value $\delta$ is represented as

$$\delta(\kappa,k)=\frac{k_{c0}k_p^\nu+ik_0(k_p^\nu+k_0^\nu)}{ik_0k_{c0}k_p^\nu\xi(0)}. \qquad (19)$$

At high frequencies the magnitude $\delta(\kappa,k)\approx k_0^\nu/(k_{c0}k_p^\nu\xi(0))$ is large. At frequencies almost up to the UV range $\delta(\kappa,k)\approx(k_{c0}+ik_0)/(ik_0k_{c0}\xi(0))\approx 1/(k_{c0}\xi(0))$.

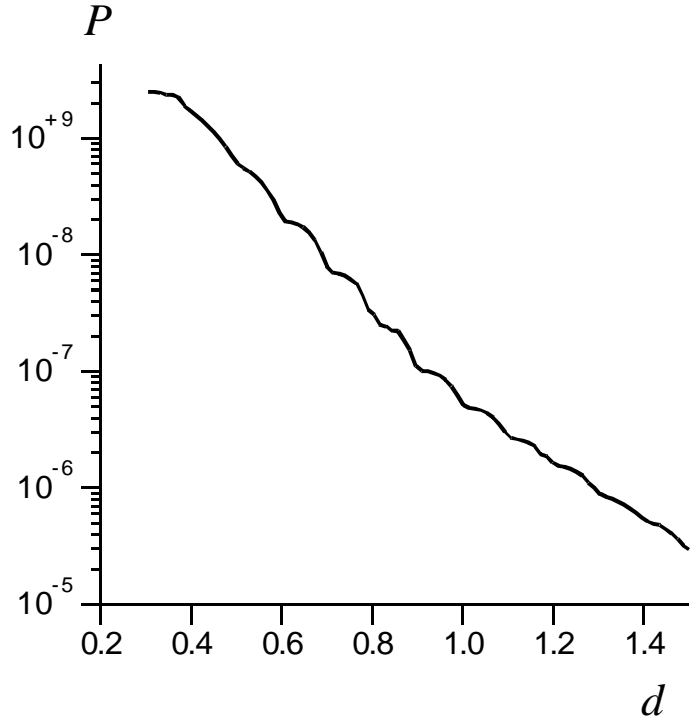

Fig. 2. Van der Waals force density $P$ (N/m$^2$) as a function of distance (nm)

The results available in the literature on models of the dispersion interaction of graphene sheets [20–27] do not adequately apply at short distances. Precisely, according to these results, as the distance decreases, the force increases indefinitely. We consider the distances to be small in the region of $1<d<\lambda_p$ nm. At $d < 1$ nm, graphene is no longer a two-dimensional continuous medium described by conduction, and at such distances, van der Waals forces between individual atoms should be considered. At the same time, we cannot talk about a distributed force (pressure). At $d=0.34$ nm, the forces of attraction and repulsion between the two sheets are compensated in alpha graphite or beta graphite, which, in addition to experiment, is confirmed by quantum calculations based on first principles models based on density functional theory. The value $d=0.34$ nm is the approximate size of the orbitals of $\pi$-electrons, and at shorter distances, electron repulsion occurs due to their mutual penetration of shells. Therefore, we assume that the model used adequately describes the interaction of graphenes at $d \geq 1$ nm distances when measured from the centers of the lattice atoms. In particular, at $d = 0$, it gives a finite attractive force, whereas in reality there should be a repulsion at distances $d \leq 0.34$ nm. The van der Waals theory should be used to calculate the forces at $d < 1$ nm. The calculation using the density functional theory method, taking into account several graphene cells, is shown in Fig. 2. and shows the finiteness of the force at $d \sim 0.34$ nm. When $d \leq 0.34$ there is a repulsion.

The Lifshitz formula is often used for graphene, replacing the reflection coefficients of the E-modes and H-modes from the half-space with the reflection coefficients from graphene [20–22,27]. The Van Kampen formula coincides with the Lifshitz formula and generalizes it to the case of multilayer flat-layered structures, including impedance sheets between layers. Therefore, both graphene sheets on a substrate (including layered ones) and weighted graphene sheets can be considered. However, in [4], an approximate formula is given for small distances, showing the law of interaction $1/d^3$, whereas at large distances it is $1/d^4$. However, the dependence $1/d^3$ corresponds to an intermediate domain and has correction terms [4]. At small distances, it is not correct, since strict calculation of integrals is necessary. Indeed, there can be no infinite force at $d \to 0$, otherwise infinite energy would be released when approaching. We show that this force is finite at $d=0$. It is known that the Lifshitz model is not applicable at distances comparable to atomic ones, which are significantly shorter than the absorption wavelengths. Therefore, the determination of dispersion forces at short distances is an urgent task [30, 31]. This is important, for example, for nanoscale diode structures [33], when the Casimir pressure together with the Coulomb force can deform a thin dielectric layer or graphene grid in resonant tunneling structures. It should be noted that the Casimir force per unit area for the interaction of almost ideal metallic half-spaces separated by a distance d also diverges as $F(d)/L^2 = \hbar c \pi^2 /(240 d^4)$ at $d \to 0$. This suggests that this formula cannot be applied to small d in the case of a gap between metal half-spaces, and its field of applicability should be determined, which is noted in particular in [4,5,34]. The real force should be obtained taking into account the dispersion in the metal [4,5,34]. Next, we show that the Casimir result for infinite conductivity is applicable only for $d \to \infty$. The aim of the work is to obtain the character of the Casimir forces in graphene at short distances. The main task is to isolate the finite part from the infinite force density as the size of the resonator increases $L \to \infty$, $D \to \infty$. When moving to continuous indexes, we get the functions

$$\alpha_e(\kappa, k, d, \delta) = -\delta \frac{k_0^2 \tan(kd/2)}{k_0^2 \delta + k \tan(kd/2)}, \tag{20}$$

$$\alpha_h(k,d,\delta) = -\delta \frac{k\tan(kd/2)}{\delta k + \tan(kd/2)}, \tag{21}$$

$$\beta_e(\kappa,k,d,\delta) = -\delta \frac{k_0^2/k}{1-\delta(k_0^2/k)\tan(kd/2)}, \tag{22}$$

$$\beta_h(k,d,\delta) = -\delta \frac{k}{1-\delta k\tan(kd/2)}, \tag{23}$$

and from (14), (15) the result for the force density associated with the emitted modes is

$$P(d,\delta) = \frac{\hbar c}{2\pi^2} \operatorname{Re} \int_0^\infty \int_0^\infty \left[\theta^{ee}(\kappa,k) + \theta^{he}(\kappa,k) + \theta^{eh}(\kappa,k) + \theta^{hh}(\kappa,k)\right] \frac{k^2\kappa}{k_0} d\kappa dk. \tag{24}$$

Here $k_0 = \sqrt{\kappa^2 + k^2}$ and

$$\theta^{ee}(\kappa,k) = \frac{2\delta(\kappa,k)kk_0^2\sin(kd) + k^2(1-\cos(kd))}{2\delta(\kappa,k)kk_0^2\sin(kd) + k^2(1-\cos(kd)) + 2\delta^2(\kappa,k)k_0^4} =$$
$$= 1 - \frac{2\delta^2(\kappa,k)k_0^4}{2\delta(\kappa,k)kk_0^2\sin(kd) + k^2(1-\cos(kd)) + 2\delta^2(\kappa,k)k_0^4}, \tag{25}$$

$$\theta^{he}(\kappa,k) = \frac{1-\cos(kd) + 2\delta k\sin(kd)}{1-\cos(kd) + 2\delta k\sin(kd) + 2\delta^2 k^2} =$$
$$= 1 - \frac{2\delta^2 k^2}{1-\cos(kd) + 2\delta k\sin(kd) + 2\delta^2 k^2}, \tag{26}$$

$$\theta^{eh}(\kappa,k) = \frac{k^2(1+\cos(kd)) - 2\delta k_0^2 k\sin(kd)}{2\delta^2 k_0^4 + k^2(1+\cos(kd)) - 2\delta k_0^2 k\sin(kd)} =$$
$$1 - \frac{2\delta^2 k_0^4}{2\delta^2 k_0^4 + k^2 + k^2\cos(kd) - 2\delta k_0^2 k\sin(kd)}, \tag{27}$$

$$\theta^{hh}(\kappa,k) = \frac{1+\cos(kd) - 2\delta k\sin(kd)}{1+\cos(kd) - 2\delta k\sin(kd) + 2\delta^2 k^2} =$$
$$= 1 - \frac{2\delta^2 k^2}{1+\cos(kd) - 2\delta k\sin(kd) + 2\delta^2 k^2}, \tag{28}$$

In the absence of graphene $\delta = \infty$ and $\theta^{\alpha\beta} = 0$, i.e., we get zero force. We have $\theta^{ee}(\kappa,k) = 0$, $\theta^{he}(\kappa,k) = 0$ at $d = 0$, and the force is determined by $\theta^{eh}(\kappa,k) = k^2/(\delta^2 k_0^4 + k^2)$ and $\theta^{hh}(\kappa,k) = 1/(1+\delta^2 k^2)$. In this case the integrals can be calculated approximately analytically by replacing the variables $\kappa = \chi\cos(\theta)$, $k = \chi\sin(\theta)$, if the angle integration is taken by the mean point theorem in $\theta = \pi/4$ and the integration domain is divided into three: $0 < \chi < k_{c0}$, $k_{c0} < \chi < k_p$ and $k_p < \chi < \infty$. Counting $\varsigma(0) < 1$ and defining in the integral

$$P(0) = \frac{\hbar c}{2\sqrt{2}\pi}\operatorname{Re}\int_0^\infty \left[\frac{1}{2\delta^2\chi^2 + 1} + \frac{1}{1+\delta^2\chi^2/2}\right]\chi^3 d\chi$$

for these regions respectively $\delta(\chi) = -i/(k_0\xi(0))$, $\delta(\kappa,k) = 1/(k_{c0}\xi(0))$ and $\delta(\kappa,k) = \chi^4/(k_{c0}k_p^4\xi(0))$, we find an approximate result

$$P(0) = \frac{13}{16\sqrt{2}\pi}\hbar c k_{c0}^2 k_p^2 \xi^2(0).$$

Note that for the imaginary $\delta$ in the first domain, the contribution is real, proportional to $k_{c0}^4$ and very small. For small $d$, the contribution from $\theta^{ee}(\kappa,k)$ and $\theta^{he}(\kappa,k)$ is proportional to $d$, and the contribution from $\theta^{eh}(\kappa,k)$ and $\theta^{hh}(\kappa,k)$ first decreases and then increases. The functions included in the square bracket in (24) vanish with large arguments, ensuring convergence. Numerical results for (24) are shown in Fig. 2. The integral was calculated in polar coordinates $\kappa = \chi\cos(\theta)$, $k = \chi\sin(\theta)$, $d\kappa dk = \chi d\chi d\theta$. At $d=0$, the force is small and attractive. At small distances $d < 1.0$ nm for $\mu_c > 1$ eV, the force is at first almost independent of the distance, then begins to increase to a maximum at distances of about 200 nm, and then decreases. For small $\mu_c < 1$ a minimum occurs at distances of the order of 0.35 nm. In the declining area, the results are close to [20], i.e. to the results of Lifshitz. To find out the result for large $d$, we replace the variables $\kappa = x/d$, $k = y/d$ and reduce the integral to the form

$$P(d,\delta) = \frac{\hbar c}{\pi^2 d^4}\text{Re}\int_0^\infty\int_0^\infty \frac{\theta^{ee}(x,y) + \theta^{he}(x,y) + \theta^{eh}(x,y) + \theta^{hh}(x,y)}{\sqrt{x^2+y^2}} y^2 x\, dx\, dy, \quad (29)$$

$$\delta(x,y,d) = \frac{k_{c0}k_p^\nu d + i\sqrt{x^2+y^2}\left(k_p^\nu + \sqrt{x^2+y^2}^\nu/d^\nu\right)}{i\sqrt{x^2+y^2}\,k_{c0}k_p^\nu \xi(0)}.$$

Consider the terms in (29). The first term has the form

$$\theta^{ee}(x,y) = \frac{2d\delta(x,y,d)y(x^2+y^2)\sin(y) + y^2 d^2(1-\cos(y))}{2d\delta(x,y,d)y(x^2+y^2)\sin(y) + y^2 d^2(1-\cos(y)) + 2\delta^2(x,y,d)(x^2+y^2)^2}.$$

Figure 3 shows the results of the contribution due to the resulting modes. The main contribution to the force in it for large d is made by small values of the integration variables, while the first oscillating term in the numerator and denominator can be neglected:

$$\theta^{ee}(x,y) \approx \frac{y^2 d^2(1-\cos(y))}{y^2 d^2(1-\cos(y)) + 2\delta^2(x,y,d)(x^2+y^2)^2}. \quad (30)$$

This value is almost independent of $d$ for small $x$ and $y$ (where the main contribution occurs) and large $d$. Therefore, its contribution to strength is also almost independent of $d$. For large $x$ and $y$, the second term in the denominator is large and ensures convergence, while the contribution $\theta^{ee}(x,y)$ is negligible. Similarly, we have

$$\theta^{eh}(x, y) = \frac{d^2 y^2(1+\cos(y)) - 2d\delta(x^2 + y^2)y\sin(y)}{d^2 y^2(1+\cos(y)) - 2d\delta(x^2 + y^2)y\sin(y) + 2\delta^2(x^2 + y^2)^2},$$

$$\theta^{he}(x, y) = \frac{d^2(1-\cos(y)) + 2d\delta y\sin(y)}{d^2(1-\cos(y)) + 2d\delta y\sin(y) + 2\delta^2 y^2},$$

$$\theta^{hh}(x, y) = \frac{d^2(1+\cos(y)) - 2d\delta y\sin(y)}{d^2(1+\cos(y)) - 2d\delta y\sin(y) + 2\delta^2 y^2},$$

and for large $d$, the main contribution to the integral, except for (30), is also given by the terms

$$\theta^{eh}(x, y) \approx \frac{d^2 y^2(1+\cos(y))}{d^2 y^2(1+\cos(y)) + 2\delta^2(x, y, d)(x^2 + y^2)^2}, \quad (31)$$

$$\theta^{he}(x, y) \approx \frac{d^2(1-\cos(y))}{d^2(1-\cos(y)) + 2\delta^2(x, y, d)y^2}, \quad (32)$$

$$\theta^{hh}(x, y) \approx \frac{d^2(1+\cos(y))}{d^2(1+\cos(y)) + 2\delta^2(x, y, d)y^2}. \quad (33)$$

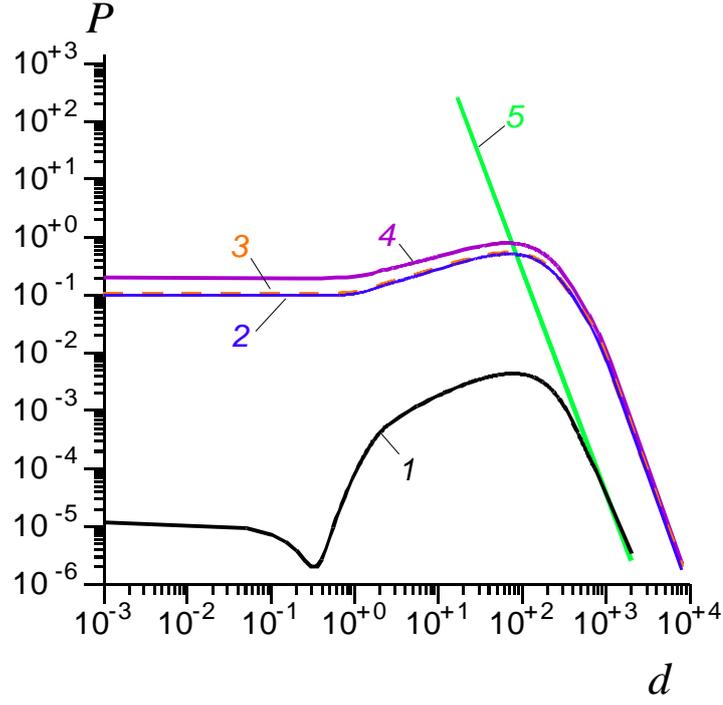

Fig. 3. The density of the Casimir attractive force $P=F/L^2$ (N/m$^2$) between two graphene sheets according to formula (14), depending on the distance $d$ (nm) at $\mu_c = 0.0783$ eV (curve 1) and $\mu_c = 7.8$ eV (curves 2–4) and different temperatures: $T=0$, (curve 2), $T=300$ K (3), $T=900$ K (4). Line 5 is the result from [20] at $T=0$. The curves are plotted for $\nu = 4$, $\omega_c = 10^{12}$ Hz

Therefore, the total contribution to the force at large $d$ is approximately proportional to $d^{-4}$. The absence of conductivity $\xi(0)=0$ (for example, at $\mu_c = 0$) leads to a lack of force. A non-zero temperature increases the conductivity, which also increases the force (Fig. 3, dashed curve 2 and curve 3). This increase is due to a change in boundary conditions. The CF has little effect on strength. Infinite conductivity also means $\delta^2(x,y,d) \to 0$ that all values (30)–(33) are equal to unity. Although the force is proportional in this case to $1/d^4$, it is infinite (integral (29) diverges), which corresponds to the results below. For large $x$ and $y$ and finite $d$ we have $\theta^{ee}(x,y) \to 0$, i.e. there is convergence, which is provided by increasing terms with $\delta^2$. This happens for such $x$, $y$, when $\delta \gg d$. An increase in $d$ in the numerical calculation of integrals requires an increase in the large upper limit of $A$. It was taken from the condition $d < 2\delta(x,y,d)$ when its doubling changes the integral by no more than 1%. The condition $d < 2\delta(x,y,d)$ allows you to evaluate the choice of the upper limit when counting with long distances: $A > d\sqrt[\nu+1]{k_{c0}k_p^\nu \xi(0)}/\sqrt{2}$. The real part of the conductivity has almost no effect on the force. The results are also weakly dependent on $\nu$. In particular, for $\nu=3$, 4, 5, and 6, we have values for curve 2 in Fig. 3 of 0.0988879, 0.0987877, and 0.099265 N/m, respectively. For $F(0)/L^2$ we have 0.132464 N/m for $\nu=5$ and 0.120154 N/m for $\nu=6$.

To account of the evanescent modes, the integral should be considered

$$P_{ev}(d,\delta) = \frac{\hbar c}{2\pi^2} \operatorname{Re} \int_0^\infty dk_0 \int_{k_0}^\infty \left[\tilde{\theta}^{ee}(\kappa,k_0) + \tilde{\theta}^{he}(\kappa,k_0) + \tilde{\theta}^{eh}(\kappa,k_0) + \tilde{\theta}^{hh}(\kappa,k_0)\right] K\kappa \, d\kappa,$$

in which the values are denoted:

$$\tilde{\theta}^{ee}(\kappa,k) = \frac{2\delta^2(k_0)k_0^4}{2\delta(k_0)Kk_0^2 \sinh(Kd) + K^2(1-\cosh(Kd)) + 2\delta^2(k_0)k_0^4},$$

$$\tilde{\theta}^{he}(\kappa,k) = \frac{2\delta^2 K^2}{1-\cos(Kd) - 2\delta K \sin K(Kd) - 2\delta^2 K^2},$$

$$\tilde{\theta}^{eh}(\kappa,k_0) = -\frac{2\delta^2 k_0^4}{2\delta^2 k_0^4 - K^2(1+\cosh(Kd)) + 2\delta k_0^2 K \sinh(Kd)},$$

$$\tilde{\theta}^{eh}(\kappa,k_0) = -\frac{\delta^2 k_0^4}{\delta^2 k_0^4 - K^2}$$

$$\tilde{\theta}^{hh}(\kappa,k_0) = \frac{2\delta^2 K^2}{1+\cosh(Kd) + 2\delta K \sinh(Kd) - 2\delta^2 K^2}.$$

$$\tilde{\theta}^{hh}(\kappa,k_0) = \frac{\delta^2 K^2}{1-\delta^2 K^2}$$

$$\tilde{\theta}^{hh}(\kappa,k_0) + \tilde{\theta}^{eh}(\kappa,k_0) = \left(-\frac{\delta^2 k_0^4}{\delta^2 k_0^4 - K^2} + \frac{\delta^2 K^2}{1-\delta^2 K^2}\right)K$$

We have subtracted from them the units obtained at $d \to \infty$ and giving an infinite and independent of $d$ contribution. In this regard, a quote from [4] is characteristic: "*Expression (2.2)* *is finite in itself, but contains terms that diverge when integrated over dω. This is the term c ω3, which occurs when integrating terms with 1/2 in curly brackets over dp. This divergent term, however, does not depend on the distance l between the bodies and therefore has no relation to the force of their mutual attraction that interests us and should be omitted. It represents the force of the reverse action of the bodies' own field on these bodies themselves, which is actually compensated by the same forces on other sides of the body.*" Now, all values for large *d*, as well as for large $\kappa$, tend to zero due to hyperbolic functions in the denominators, which corresponds to exponential attenuation of plasmons. The results of the contribution of evanescent modes to the force are shown in Fig. 4, Curve 2. The contributions of the resulting modes (1,3,4) are also shown there.

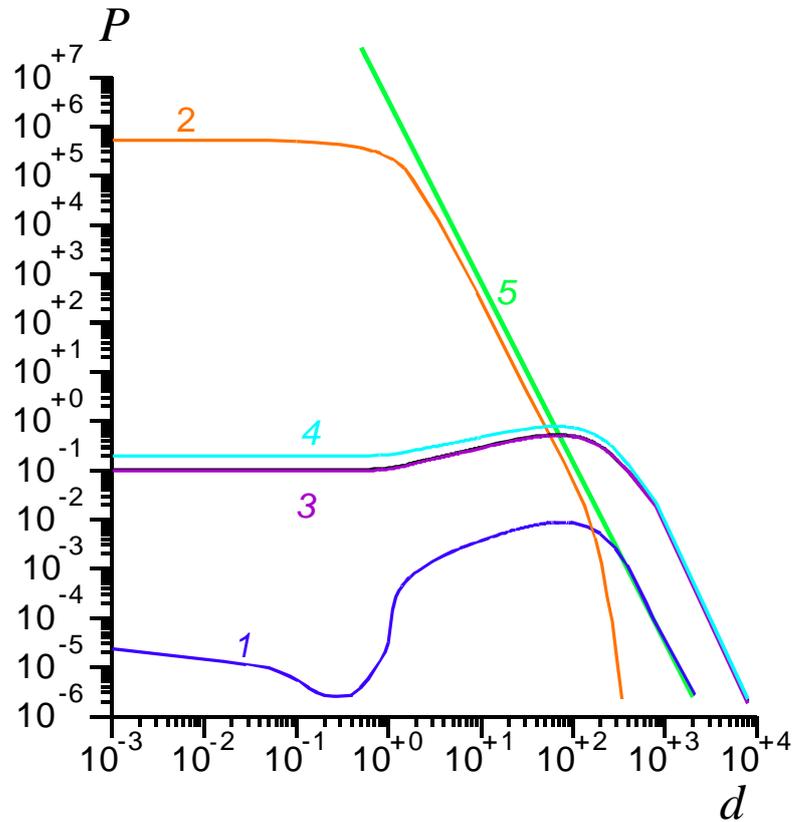

Fig. 4. Densities (N/m$^2$) of the attractive force $\tilde{P}$ (curves 1,3,4) and $P_{ev}$ (2) between two graphene sheets as a function of the distance *d* (nm) at $\mu_c = 0.0783$ eV (curves 1,2) and $\mu_c = 7.8$ eV (curves 3,4). Curves 1–3 are plotted at a temperature of *T*=600, curve 4 is at *T*=900 (K). Line 5 is the result from [20] at *T*=0. Everywhere $\nu = 4$, $\omega_c = 10^{12}$ Hz

Let us consider another summation method in (1) proposed by Van Kampen et al. [35]. It is also based on the introduction of characteristic equations using the theorem (principle) of the argument and has been considered in a number of papers, for example, in [29,31]. According to [29,31,35], summation (1) for infinite resonator sizes defines the Casimir pressure as $P(d) = -\partial_d E(d) = P_e(d) + P_h(d)$, where

$$P_{e,h}(d) = -\frac{\hbar}{4\pi^2} \int_0^\infty \kappa d\kappa \int_0^{i\infty} d\omega \frac{\sqrt{\kappa^2 - (\omega/c)^2}}{f_{e,h}(\kappa,\omega,d)} =$$
$$= -\frac{\hbar c}{4\pi^2} \int_0^\infty \kappa d\kappa \int_0^\infty dk \frac{K}{f_{e,h}(\kappa,k,d)}. \tag{34}$$

This pressure inside the graphene sheets differs in sign from (13). For the structure of Fig. 1 at $D = \infty$, it is possible to write the characteristic equations obtained by the impedance transformation method [36,37] in the form $f_{e,h}(k_0, k_z) = 0$, where $k_z = \sqrt{k_0^2 - \kappa^2}$, $\kappa^2 = k_x^2 + k_y^2$, $y_e = 1/y_h = k_0/k_z = k/K$, and

$$-f_{e,h}(k_0, k_z d) = \varsigma \rho_{e,h} \frac{1 + i\varsigma \rho_{e,h} \tan(k_z d)}{1 + i(1 + \varsigma \rho_{e,h})\tan(k_z d)} + 1. \tag{35}$$

The transfer matrix method leads to the same equation. Strictly speaking, these are the same equations (2), but with $D \to \infty$ and $k_z = -iK$. Using the imaginary frequency $\xi = i\omega$, we have $k_z = -iK = -i\sqrt{\kappa^2 + k^2}$, $k = \xi/c = ik_0$, $\rho_e = 1/\rho_h = K/k$, $\varsigma(\xi) = \varsigma(0)/(1 + \xi/\omega_c(\xi))$, $y_{e,h} = 1/\rho_{e,h}$, and

$$-f_{e,h}(\kappa, k, d) = \frac{\varsigma(k)\rho_{e,h} + (\varsigma(k)\rho_{e,h})^2 \tanh(Kd)}{1 + (1 + \varsigma(k)\rho_{e,h})\tanh(Kd)} + 1.$$

For any $d$, this function gives a divergent result when used in (34). In the absence of graphene sheets $f_{e,h}(\kappa, k, d) = -1$, the result also diverges. Since value $f_{e,h}(\kappa, k, \infty)$ does not depend on $d$, according to [29], terms that do not depend on $d$ (due to the infinite vacuum energy) should be subtracted from result (34) and taken

$$\tilde{P}_{e,h}(d) = -\frac{\hbar c}{4\pi^2} \int_0^\infty \int_0^\infty K\left(\frac{1}{f_{e,h}(\kappa,k,d)} - \frac{1}{f_{e,h}(\kappa,k,\infty)}\right) dk \kappa d\kappa. \tag{36}$$

Since $\tanh(Kd) = 1$ for infinitely large $d$, we obtain the value of the function

$$f_{e,h}(\kappa, k, \infty) = -\frac{2 + 2\varsigma(k)\rho_{e,h} + (\varsigma(k)\rho_{e,h})^2}{2 + \varsigma(k)\rho_{e,h}}.$$

We have $f_{e,h}(\kappa, \infty, d) = -1$, $f_{e,h}(\kappa, \infty, \infty) = -1$, which ensures convergence. However, you may lose the multiplier. The sign in (35) is taken so that for large $d$ the result corresponds to other characteristic equations. With a very high conductivity $f_{e,h}(\kappa, k, d) \approx f_{e,h}(\kappa, k, \infty) \approx \varsigma(k)\rho_{e,h}$, and

(36) gives uncertainty. Expressing $\tanh(Kd)$ through $\exp(2Kd)$, we get $f_{e,h}(\kappa,k,d)\tilde{Q}_{e,h} = \tilde{f}_{e,h}(\kappa,k,d)$ where

$$\tilde{f}_{e,h}(\kappa,k,d) = \frac{1+\varsigma(k)\rho_{e,h}+\varsigma^2(k)\rho_{e,h}^2}{\varsigma^2(k)\rho_{e,h}^2}\exp(2Kd)-1,$$

$$\tilde{Q}_{e,h} = -\frac{(1+\varsigma(k)\rho_{e,h})\exp(2Kd)-\varsigma(k)\rho_{e,h}}{\varsigma^2(k)\rho_{e,h}^2},$$

in other words, the function $\tilde{f}_{e,h}(\kappa,k,d)$ corresponds to the original one up to a multiplier $\tilde{Q}_{e,h}$. Using the stitching (mode matching) method from [29], we find

$$\tilde{\tilde{f}}_{e,h}(\kappa,k,d) = \frac{(2+\varsigma\rho_{e,h})^2}{\varsigma^2\rho_{e,h}^2}\exp(2Kd)-1. \tag{37}$$

All functions define frequency spectra or plasmon spectra, which at $D \to \infty$ are infinitely close and become continuous. In particular, for $d=0$, all functions give $\varsigma\rho_{e,h}+1=0$. For $\varsigma \to \infty$ we have $\tilde{\tilde{f}}_{e,h}(\kappa,k,d) = \tilde{f}_{e,h}(\kappa,k,d) = \exp(2Kd)-1$, and both functions correspond to the Casimir problem [29], so the multipliers for them are chosen correctly. When $d \to 0$, both functions In (34) give a divergent result (due to the contribution of electric modes), which is explained by the absence of a decrease in this case on the infinite semicircle of the complex frequency plane. At small distances, the convergence is worse, the smaller it is. At the same time, for $\varsigma \to \infty$ we have $f_{e,h}(\kappa,k,d) \to \infty$, $\tilde{Q}_{e,h} \to 0$. Therefore, for small $d$, it is advisable to use the result (36) with a correction factor. For the Casimir problem $\varsigma \to \infty$, it gives uncertainty. Consider (36) for $d \to \infty$. We have $\tanh(Kd) \approx 1-2\exp(-2Kd)$ and

$$\frac{1}{f_{e,h}(\kappa,k,d)} - \frac{1}{f_{e,h}(\kappa,k,\infty)} = 2\varphi_{e,h}(\kappa,k)\exp(-2Kd),$$

$$\varphi_{e,h}(\kappa,k) = \frac{1+\varsigma\rho_{e,h}}{2+2\varsigma\rho_{e,h}+(\varsigma\rho_{e,h})^2} + \\ + \frac{(2+\varsigma\rho_{e,h})[1+\varsigma\rho_{e,h}+(\varsigma\rho_{e,h})^2]}{[1+\varsigma\rho_{e,h}+(\varsigma\rho_{e,h})^2]^2}.$$

With high conductivity $\varsigma$ we have $\varphi_{e,h}(\kappa,k) = 2/(\varsigma\rho_{e,h})$. At low one, it is convenient to take this function in the form

$$\varphi_{e,h}(\kappa,k) = 1 - \frac{(\varsigma\rho_{e,h})^2/2}{2+2\varsigma\rho_{e,h}+(\varsigma\rho_{e,h})^2} + $$

$$-\frac{\varsigma\rho_{e,h}+(\varsigma\rho_{e,h})^2+(\varsigma\rho_{e,h})^3+(\varsigma\rho_{e,h})^4/2}{(2+2\varsigma\rho_{e,h}+(\varsigma\rho_{e,h})^2)^2}.$$

Here, the contribution is mainly made by low frequencies, so you can take $\varsigma(0)$. Now it is quite simple to show the dependence $1/d^4$ for strength. We make substitutions $2Kd = w$, $kd = v$, and we get

$$\tilde{P}_{e,h} = -\frac{\hbar c}{16\pi^2 d^4}\int_0^\infty dv \int_{2v}^\infty \exp(-w)\varphi_{e,h}(w,v)w^2 dw.$$

If the conductivity is extremely low, then $\varphi_{e,h}(w,v) \approx 1$, and we have $\tilde{P}_{e,h} = -3\hbar c/(16\pi^2 d^4)$. With low conductivity, taking the decomposition $\varphi_{e,h}(w,v) \approx 1 - \varsigma(0)\rho_{e,h}/4$, we find

$$\tilde{P}_h = -\frac{\hbar c}{16\pi^2 d^4}\left(3 - \frac{\varsigma(0)}{8}\right),$$

but $\tilde{P}_e$ diverges logarithmically at low frequencies. Divergence is related to decomposition. If we take the lower limit $v = k_{c0}d$, then at the distances $d \sim 1/k_{c0}$ will be

$$\tilde{P}_e \approx -\frac{\hbar c}{16\pi^2 d^4}\left(3 - \frac{5\varsigma(0)}{2}\right).$$

The final result is obtained by integrating without using decomposition. A similar result with a specific value of the coefficient at $1/d^4$ can be obtained by integration in parts. Therefore, using (36) for large $d$ yields a decrease according to the law $1/d^4$ (Fig. 3, 4). For the result (36) $\tilde{P}_{e,h}(d) \to 0$ for $d \to \infty$, as well as for $\varsigma \to 0$, since $f_e(\kappa,k,d) \to -1$ and $f_e(\kappa,k,\infty) \to -1$. It sums up all the frequencies because on a large semicircle in the complex frequency plane $1/f_{e,h}(\kappa,k,d) - 1/f_{e,h}(\kappa,k,\infty) \to 0$. Consider the case of $d=0$. It is necessary to calculate the integrals

$$\tilde{P}_e(0) = -\frac{\hbar c}{4\pi^2}\int_0^\infty\int_0^\infty Kk\left(\frac{2k+\varsigma(k)K}{2k^2+2\varsigma(k)Kk+\varsigma^2(k)K^2} - \frac{1}{\varsigma(k)K+k}\right)dk\kappa d\kappa,$$

$$\tilde{P}_h(0) = -\frac{\hbar c}{4\pi^2}\int_0^\infty\int_0^\infty K^2\left(\frac{2K+\varsigma(k)k}{2K^2+2\varsigma(k)kK+\varsigma^2(k)k^2} - \frac{1}{\varsigma(k)k+K}\right)dk\kappa d\kappa.$$

Making the substitution $\kappa = \chi\cos(\theta)$, $k = \chi\sin(\theta)$, $d\kappa dk = \chi d\chi d\theta$, $\varsigma(\chi,\theta) = \varsigma(0)k_{c0}k_p^4/(k_{c0}k_p^4 + k_p^4\chi\sin(\theta) + \chi^5\sin^5(\theta))$, $K = \chi$, we have

$$\tilde{P}_e(0) = \frac{\hbar c}{4\pi^2} \int_0^\infty \chi^3 d\chi \int_0^{\pi/2} \left( \frac{1}{\varsigma + \sin(\theta)} - \frac{\sin(\theta) + \varsigma/2}{\sin^2(\theta) + \varsigma \sin(\theta) + \varsigma^2/2} \right) \sin(\theta)\cos(\theta) d\theta,$$

$$\tilde{P}_h(0) = \frac{\hbar c}{4\pi^2} \int_0^\infty \chi^3 d\chi \int_0^{\pi/2} \left( \frac{1}{\varsigma \sin(\theta) + 1} - \frac{1 + \varsigma \sin(\theta)/2}{1 + \varsigma \sin(\theta) + \varsigma^2 \sin^2(\theta)/2} \right) \cos(\theta) d\theta.$$

We have zero results at $\varsigma = 0$. When $\theta = 0$ the parentheses turn to zero. At any $\theta \neq 0$ the value $\varsigma(\chi, \theta) \to 0$ at $\chi \to \infty$, ensuring convergence. Since the dependence on the angle is not strong, the angle integral can be calculated approximately by the mean value theorem, taking it at the midpoint $\theta = \pi/4$:

$$\tilde{P}_e(0) = \frac{\hbar c}{16\pi} \int_0^\infty \left( \frac{1}{\varsigma(\chi) + 1/\sqrt{2}} - \frac{\sqrt{2} + \varsigma(\chi)}{1 + \sqrt{2}\varsigma(\chi) + \varsigma^2(\chi)} \right) \chi^3 d\chi,$$

$$\tilde{P}_h(0) = \frac{\hbar c}{8\sqrt{2}\pi} \int_0^\infty \left( \frac{1}{\varsigma(\chi)/\sqrt{2} + 1} - \frac{1 + \varsigma/2\sqrt{2}}{1 + \varsigma/\sqrt{2} + \varsigma^2/4} \right) \chi^3 d\chi.$$

Here $\varsigma(\chi) = \varsigma(0) k_{c0} k_p^4 / \left( k_{c0} k_p^4 + k_p^4 \chi/\sqrt{2} + \chi^5/(4\sqrt{2}) \right)$. For the area $0 < \chi < k_{c0}$ the brackets can be removed from under the integral at $\varsigma(\chi) = \varsigma(0)$, and the integrals give a small multiplier $k_{c0}^4/4$. For large $\chi$, the function $\varsigma(\chi) \approx 4\sqrt{2}\varsigma(0) k_{c0} k_p^4 / \chi^5$ is small in area $\chi > \sqrt[5]{k_{c0} k_p^4}$. Let us consider the case of small $k_{c0}$ and small value $\varsigma^2(0)$, when $\varsigma(\chi) = \varsigma(0) k_{c0} k_p^4 / \left( k_p^4 \chi/\sqrt{2} + \chi^5/(4\sqrt{2}) \right)$ and

$$\tilde{P}_e(0) \approx -\frac{\hbar c \varsigma(0) k_{c0} k_p^4}{2\sqrt{2}\pi} \int_{k_c}^\infty \frac{\chi^3 d\chi}{\chi^5 + 4k_p^4 \chi + 8\varsigma(0) k_{c0} k_p^4},$$

$$\tilde{P}_h(0) \approx -\frac{\hbar c \varsigma(0) k_{c0} k_p^4}{4\sqrt{2}\pi} \int_{k_c}^\infty \frac{\chi^3 d\chi}{\chi^5 + 4k_p^4 \chi + 4\varsigma(0) k_{c0} k_p^4}.$$

Integrals are taken, but it is quite difficult. This result is proportional to the CF. Neglecting the small third small term in the denominators, we have $\tilde{P}_e(0) = 2\tilde{P}_h(0)$ and $\tilde{P}(0) = -3\varsigma_0 \mu_c k_p^3/(4\sqrt{2}\pi)$, or $\tilde{P}(0) = -3\varsigma_0 \mu_c k_p^3/(4\sqrt{2}\pi)$. At $\tilde{P}(0) = -3\varsigma_0 \mu_c k_p^3/(4\sqrt{2}\pi)$ eV, we get $\tilde{P}(0) \sim -2 \cdot 10^8$ N/m$^2$.

Consider the result (34) with the function (37) $\tilde{\tilde{f}}_{e,h}$:

$$P_{e,h}(d) = -\frac{\hbar c}{4\pi^2} \int_0^\infty \kappa d\kappa \int_0^\infty dk \frac{K \varsigma^2 \rho_{e,h}^2}{(2 + \varsigma \rho_{e,h})^2 \exp(2Kd) - \varsigma^2 \rho_{e,h}^2}.$$

At $d=0$ and high frequencies, the convergence is provided by the decreasing function $\varsigma^2$, but $P_e(0)$ diverges at low frequencies. By making the substitution $u = K = \sqrt{\kappa^2 + k^2}$, $udu = K = \kappa d\kappa$, we obtain the integrals

$$P_{e,h}(d) = -\frac{\hbar c}{4\pi^2} \int_0^\infty dk \int_k^\infty du \frac{\varsigma^2(k)}{(2+\varsigma\rho_{e,h})^2} \frac{\rho_{e,h}^2 u^2 \exp(-2ud)}{1-\exp(-2ud)\varsigma^2(k)\rho_{e,h}^2/(2+\varsigma\rho_{e,h})^2},$$

which are easy to calculate by integration in parts. The second fraction has either a third-order zero in $u$, or a second-order zero in $k$ and a first-order zero in $u$. After integration in parts of u and substitution, we obtain a multiplier function with a second-order zero in $k$. Triple integration in parts with respect to $k$ leaves a term of order $1/d^4$ and a more strongly decreasing integral. The decomposition can be continued. Let's now consider the replacement $\kappa^2 = k^2(p^2-1)$, $\kappa d\kappa = k^2 p dp$, $K = kp$, $k = y/(2pd)$, $dk = dy/(2pd)$:

$$P_{e,h}(d) = -\frac{\hbar c}{64\pi^2 d^4} \int_1^\infty p^{-2} dp \int_0^\infty \frac{\gamma_{e,h} y^3 \exp(-y) dy}{1-\gamma_{e,h} \exp(-y)} \approx$$

$$\approx -\frac{\hbar c}{64\pi^2 d^4} \int_1^\infty p^{-2} dp \int_0^\infty \gamma_{e,h} y^3 (\exp(-y) + \gamma_{e,h} \exp(-2y) + \ldots) dy,$$

$$\gamma_e = \frac{\varsigma^2(y,d)p^2}{(2+\varsigma(y,d)p)^2}, \quad \gamma_h = \frac{\varsigma^2(y,d)/p^2}{(2+\varsigma(y,d)/p)^2}.$$

Here $\varsigma(y,p) = \varsigma(0)k_{c0}k_p^4 / (k_{c0}k_p^4 + k_p^4 y/(2pd) + y^5/(2pd)^5)$, and in the decomposition, we used the condition $\gamma_{e,h} < 1$. With a large $d$ you can take $\varsigma(y,p) = \varsigma(0)$. Then these integrals can be calculated by integration in parts over $y$, while the presence of a third-order zero nullifies the first three substitutions. As a result, we get a power decomposition in $1/d^n$, $n \geq 4$. However, they are easier to calculate numerically. Going to the limit $d \to \infty$ means $\varsigma(y,p) = \varsigma(0)$. With a very small value $\varsigma(0)$, you can take $\gamma_h \approx \varsigma^2(0)/(4p^2)$, and then

$$P_h(d) \approx -\frac{\hbar c \varsigma^2(0)}{128\pi^2 d^4}.$$

In this case $P_e(d)$ should be calculated by taking $\gamma_e = \varsigma^2(0)p^2/(2+\varsigma(0)p)^2$:

$$P_e(d) = -\frac{3\hbar c \varsigma(0)}{16\pi^2 (2+\varsigma(0)) d^4}.$$

With low conductivity, then we get $P(d) \approx -3\hbar c \varsigma(0)/(32\pi^2 d^4)$ what is different from the result $\tilde{P}(d) = -3\hbar c/(8\pi^2 d^4)$. Since the latter is independent of $\varsigma(0)$, it should be considered less reliable. For very small distances $d \ll 1/\sqrt[5]{k_{c0}k_p^4}$, it is necessary to take $\varsigma(y,p) = \varsigma(0)(2p)^5 d^5 k_{c0} k_p^4 / y^5$, and then, after integration the terms proportional to positive powers of $d$ arise in $P(d)$. The result is quite difficult to obtain, and with small $d$, the result $P(d)$ should be considered less reliable than $\tilde{P}(d)$. If it is extremely large $\varsigma(0) \to \infty$, it should be taken $\gamma_e = \gamma_h = 1$, and then we get the Casimir result. Consider the result (34) with the function $\tilde{f}_{e,h}$:

$$\overline{P}_{e,h}(d) = -\frac{\hbar c}{4\pi^2} \int_0^\infty \kappa d\kappa \int_0^\infty dk \frac{K\varsigma^2 \rho_{e,h}^2}{(1 + \varsigma(k)\rho_{e,h} + \varsigma^2(k)\rho_{e,h}^2)\exp(2Kd) - \varsigma^2 \rho_{e,h}^2}.$$

With low conductivity, if we ignore the value $\varsigma^2(k)\rho_{e,h}^2$ in parenthesis, it is four times greater than $P(d)$. With a very high conductivity and a large distance $\varphi_{e,h} = 2/(\varsigma(0)\rho_{e,h})$, therefore, in order to match the result $\tilde{P}_{e,h}(d)$ to $P_{e,h}(d)$ over long distances, the function should be multiplied by $Q_{e,h} = \varsigma(0)\rho_{e,h}/4$. For very low conductivity, and for matching over long distances, the function should be multiplied by $Q_{e,h} = \varsigma(0)\rho_{e,h}/4$. For very low conductivity $\varphi_{e,h} = 1$, and for matching over long distances the function $f_{e,h}$ should be multiplied by $Q_{e,h} = \varsigma^2(0)\rho_{e,h}^2/8$.

For a fixed transverse wavenumber $\kappa$, equations of type (35) determine the resonant frequencies of a Fabry-Perrault resonator with permeable screens, and for a dependence, plasmons along graphene sheets $\kappa(k_0)$. The resonances correspond to fast flowing plasmons. When $d=0$ two sheets are connected and have double the conductivity $2\varsigma$, and it follows $\varsigma\rho_{e,h} + 1 = 0$. These are the dispersion equations $\kappa = k_0\sqrt{1-1/\varsigma^2}$ and $\kappa = k_0\sqrt{1-\varsigma^2}$ for plasmons in a graphene bilayer with double conductivity [38]. In the general case $f_{e,h}(k_0, k_z) = 0$ and $\tilde{\tilde{f}}_{e,h}(k_0, k_z) = 0$ are the equations for E- and H-plasmons with electric and magnetic walls between the sheets [36, 37]. The function $f_{e,h}(\kappa, k, d)$ can be divided into two multipliers [37]: $f_{e,h}(\kappa, k, d) = f_{e,h}^+(\kappa, k, d) f_{e,h}^-(\kappa, k, d)$, passing to $\tanh(Kd/2)$ for the tangent half of argument. Exactly $\tilde{f}_{e,h}(\kappa, k, 0) = 4(1 + \varsigma\rho_{e,h})/(\varsigma\rho_e)^2$ where the plus corresponds to the magnetic wall and the minus corresponds to the electric wall, which determines the even and odd plasmons. At $d=0$ $\tilde{f}_{e,h}(\kappa, k, 0) = 4(1 + \varsigma\rho_{e,h})/(\varsigma\rho_e)^2$, and integral (34) converges due to decreasing $\varsigma$ with frequency. When $\varsigma \to \infty$ we obtain the result of Casimir $P(d) = \hbar\pi^2 c/(240 d^4)$ [29]. In the general case, it is possible to obtain a power decomposition $1/d^n$. The function $\tilde{f}_{e,h}(\kappa,k,d)$ at $\varsigma \to \infty$ also gives the Casimir result. The equation $\tilde{\tilde{f}}_{e,h}(\kappa,k,d) = 0$ also splits into two for an even and an odd plasmon [39]

$$\tilde{\tilde{f}}^{\pm}_{e,h}(\kappa,k,d) = \frac{2+\varsigma\rho_{e,h}}{\varsigma\rho_{e,h}}\exp(Kd) \pm 1 = 0,$$

with $\quad \tilde{\tilde{f}}_{e,h}(\kappa,k,d) = f^+_{e,h}(\kappa,k,d) f^-_{e,h}(\kappa,k,d).\quad$ After the substitution $\exp(Kd) = (1+\tanh(Kd/2))/(1-\tanh(Kd/2))$ we have $\bar{f}^+_{e,h}(\kappa,k,d) = 1 + \varsigma\rho_{e,h} + \tanh(Kd/2)$, $\bar{f}^-_{e,h}(\kappa,k,d) = (1+\varsigma\rho_{e,h})^{-1} + \tanh(Kd/2)$, or $\tanh(Kd/2) + (1+\varsigma\rho_{e,h})^{\pm 1} = 0$. The plus corresponds to the case of even oscillations [39], the minus corresponds to the electric wall. Such transformations do not change the roots, but they change the amplitudes of the functions, which is important to take into account in the Van Kampen method. All functions at $d=0$ give roots $\rho_{e,h}\varsigma + 1 = 0$. For $\varsigma = \infty$ all functions with an exponent give the Casimir model. However, they have different values for $\varsigma = 0$ and $d = 0$, which corresponds to different multipliers.

The possibility of using dissipative structures in calculations of the Casimir force is discussed in a number of papers [29–31,40]. The plasma model without dissipation is mainly used, although the Drude model is also used [30]. In this case, the imaginary parts are discarded. Formulas (34) and (36) do not contain imaginary parts. The Casimir result with infinite conductivity also contains no imaginary parts and is directly obtained from (34). The result (24) also contains no imaginary parts at $\omega_{c0} \to 0$. Note that the Lifshitz formula [1] also takes the real part, and the Van Kampen method accurately gives the Lifshitz and Casimir formulas [29,35].

To clarify the limits of applicability of the result [1], consider summation, when two ideal sheets have infinite conductivity: $k_{zl} = l\pi/d$ and

$$F = \frac{\hbar}{2}\partial_d \sum_{mnl}(\omega^e_{mnl} + \omega^h_{mnl}) = -\frac{\hbar c \pi^2}{d^3}\sum_{mnl}\frac{l^2}{\sqrt{k^2_{xm} + k^2_{ym} + k^2_{zl}}}. \quad (38)$$

Taking into account the degeneration of the oscillations, we doubled the result. This value diverges even if we consider the force density. Replacing the double sum with an integral, we find

$$\frac{F}{L^2} = \frac{\hbar}{2}\partial_d \sum_{mnl}(\omega^e_{mnl} + \omega^h_{mnl}) = -\frac{\hbar c \pi^4}{2d^4}\sum_{l=1}l^2\left(\sqrt{d^2\kappa^2_{\max}/\pi^2 + l^2} - l\right). \quad (39)$$

The integral diverges, so we have limited the upper limit. The series also diverges even if $\kappa_{max}^2$ is finit. This suggests that the "finite value" of (38) or (39) can be distinguished only in the limit $d \to \infty$, i.e. the Casimir result is strictly applicable for large $d$, or rather for $d \to \infty$. When $l^2 >> \alpha^2 = d^2\kappa_{max}^2/\pi^2$ we get the remainder of the series

$$\sum_{l>>[\alpha]}\left(\frac{1}{2}\alpha^2 l - \frac{1}{2}\alpha^4/l + ...\right) \approx \frac{\alpha^2}{4}\left(l_{max}^2 - \alpha^2\right) \approx \frac{\alpha^2}{4}l_{max}^2.$$

It has the dependence $d^{-2}$. Let's take $l_{max} = 10\alpha$ and get

$$\frac{F}{L^2} = -\frac{\hbar c \pi^4}{2d^4}\sum_{l=1}^{10[\alpha]} l^2\left(\sqrt{\alpha^2+l^2}-l\right) - \frac{\hbar c \pi^2 \kappa_{max}^2 l_{max}^2}{8d^2}.$$

For small $d$ such that $d^2\kappa_{max}^2/\pi^2 < 1$, we have

$$\sqrt{d^2\kappa_{max}^2/\pi^2 + l^2} - l = \left(\frac{1}{2}\frac{\alpha^2}{l} - \frac{1}{8}\left(\frac{\alpha^4}{l^3}\right) + \frac{3}{48}\frac{\alpha^6}{l^5} - \frac{15}{384}\frac{\alpha^8}{l^7} + ....\right). \tag{40}$$

The series (39) diverges due to the first and second terms in (40). If we limit ourselves to the maximum index $l_{max}$ and number $\kappa_{max}$, we get

$$\frac{F}{L^2} = -\frac{\hbar c \pi^4 \kappa_{max}^2}{24\pi^2}\left(\frac{l_{max}^3}{d^2} - \frac{1}{8}\frac{\kappa_{max}^2}{\pi^2}(C+\psi(l_{max}+1)) + \frac{3\varsigma(3)}{48}\frac{d^2\kappa_{max}^4}{\pi^4} - \frac{15\varsigma(5)}{384}\frac{d^4\kappa_{max}^6}{\pi^6} + ...\right). \tag{41}$$

The second infinite term in (41) is independent of $d$. At short distances, the force becomes proportional to $d^{-2}$, but the series (41) diverges at $\kappa_{max} \to \infty$. At very small distances, according to (39) at $d \to 0$, we have uncertainty $0/0$. Opening it according to Lopital, we have

$$\lim_{d\to 0}\frac{\kappa_{max}^2}{3\pi^2 d^2}\sum_{l=1}\frac{l^2}{\sqrt{d^2\kappa_{max}^2/\pi^2+l^2}} = ... = -\frac{\kappa_{max}^4}{\pi^4}\sum_{l=1}l^{-1} = -\infty,$$

which indicates the logarithmic divergence of the coefficient at $d^{-2}$. Infinite force is a consequence of infinite conduction and infinite vacuum energy. For finite two-dimensional conductivity, the force for physical reasons must be finite at $d=0$. For large $d$ it can be considered $d^2\kappa_{max}^2/\pi^2 >> l_{max}^2$. Then we get

$$\frac{F}{L^2} = -\frac{\hbar c \pi^4}{2d^4}\sum_{l=1}\left(\frac{l_{max}^3 d\kappa_{max}}{3\pi}\alpha + \frac{1}{2}\frac{\pi d_{max}^3}{5d\kappa_{max}} - \frac{1}{8}\frac{\pi^3 l_{max}^7}{7d^3\kappa_{max}^3 \alpha^3} + \frac{3}{48}\frac{l_{max}^8}{8\alpha^5} - ... - \frac{l_{max}^4}{4}\right).$$

The question arises about the Casimir attractive force $F(d)/L^2 = \hbar c \pi^2/(240d^4)$. To do this, let's consider how it was obtained in [1]. Having examined a resonator of size $d$, Casimir found its energy

$$\frac{\hbar}{2}\sum \omega_{mnl} = \frac{\hbar c L^2}{2\pi} \sum_l \int_0^\infty \sqrt{\kappa^2 + (l\pi/d)^2}\, \kappa d\kappa. \qquad (42)$$

From this infinite energy, he subtracted the same infinite energy, but with a large $d$, then moving to the limit $d \to \infty$ and replacing the summation with integration, i.e. he subtracted the integral [1]

$$\frac{d}{\pi}\int_0^\infty \sqrt{k_z^2 + \kappa^2}\, \kappa d\kappa \int_0^\infty dk_z. \qquad (43)$$

The goal was to subtract the infinite energy of the vacuum. Casimir's result has the form [1]

$$\delta \frac{\hbar}{2}\sum \omega_{mnl} = \frac{\hbar c L^2}{2\pi}\left[\sum_l \int_0^\infty \sqrt{\kappa^2 + (l\pi/d)^2}\,\kappa d\kappa - \frac{d}{\pi}\int_0^\infty \sqrt{k_z^2 + \kappa^2}\,\kappa d\kappa \int_0^\infty dk_z\right]. \qquad (44)$$

There is a possibility of incorrect interpretation here. In formula (42) and in the first term in (44), it is as if $d$ is any, including small, and in (43) and in the second integral in (44) it is extremely large. The extremely large $d$ corresponds to the expansion of the sheets to infinity, i.e., taking into account the infinite energy of empty space. Therefore, everywhere in (44) $d$ must be extremely large. Formula (42) gives energy inside a resonator of size $d$ with perfectly conductive sheets, which fully corresponds to the Casimir model. Such sheets in the $D$-size resonator (Fig. 1) discussed above are under pressure from the energy density in it outside the sheets, which is greater than between the sheets. Since the resonator is divided into two parts, the energy of both parts should be taken. Their size $(D-d)/2$. Therefore, instead of (44) in our case, we should take

$$\delta \frac{\hbar}{2}\sum \omega_{mnl} = \frac{\hbar c L^2}{2\pi}\sum_l \int_0^\infty \left(\sqrt{\kappa^2 + (l\pi/d)^2} - 2\int_0^\infty \sqrt{\kappa^2 + (2l\pi/(D-d))^2}\right)\kappa d\kappa, \qquad (45)$$

counting $D \to \infty$. For the right integral $D \to \infty$ in the limit of large $D$ has continuous values. For the left integral, one cannot assume $(l\pi/d)^2 = k_{zl}^2$. Replacing the sum with the integral, we get

$$2\sum_l \sqrt{\kappa^2 + (2l\pi/D - d)^2} \approx \frac{(D-d)}{\pi}\int_0^{k_{max}} \sqrt{\kappa^2 + k_z^2}\, dk_z =$$

$$= \frac{(D-d)}{\pi}\left[\frac{k_{max}\sqrt{\kappa^2 + k_{max}^2}}{2} + \frac{\kappa^2}{2}\ln\left(\frac{k_{max} + \sqrt{\kappa^2 + k_{max}^2}}{\kappa}\right)\right].$$

If we consider $\kappa \ll k_{max}$ in this result, then

$$2\sum_l \sqrt{\kappa^2+(2l\pi/D-d)^2} \approx \frac{(D-d)}{\pi}\int_0^{k_{max}}\sqrt{\kappa^2+k_z^2}dk_z =$$
$$=-\frac{d}{\pi}\left[k_{max}\sqrt{\kappa^2+k_{max}^2}+\frac{\kappa^2}{2}\ln\left(\frac{k_{max}+\sqrt{\kappa^2+k_{max}^2}}{\kappa}\right)\right] \quad . \tag{46}$$

Here we have omitted the infinite proportional to $D$ and the $d$-independent term. Differentiating the infinite contribution (46), we see that it gives a constant force. Now we have a difference from [1]:

$$\delta\frac{\hbar}{2}\sum \omega_{mnl} = \frac{\hbar c L^2}{2\pi}\left\{\sum_l \int_0^\infty \sqrt{\kappa^2+l^2(\pi/d)^2}\kappa d\kappa + \frac{d}{\pi}\int_0^\infty\int_0^{k_{max}}\sqrt{\kappa^2+k_z^2}dk_z \kappa d\kappa\right\}.$$

It is essential that $k_z^2 \neq l^2(\pi/d)^2$. Making a Casimir substitution $u = d^2\kappa^2/\pi^2$, we get

$$\delta\frac{\hbar}{2}\sum \omega_{mnl} = \frac{\hbar c \pi^2 L^2}{4d^3}\left\{\sum_l \int_0^\infty \sqrt{u+l^2}du + \frac{d}{D}\int_0^\infty\int_0^\infty \sqrt{u+\frac{d^2}{D^2}l^2}dldu\right\}.$$

After dividing by $L^2$ we can assume that the pressure associated with the right integral is zero. There are different functions here, and the Euler-Maclaurin formula cannot be applied. In a large resonator, the double integral is reset to zero. Thus, formula (39) correctly describes the Casimir pressure. It's endless for perfect sheets. But there are no perfect sheets in nature. Kazimir introduced a clipping multiplier $f(u,l)$ that can be associated with a real metal, for which at low frequencies one can consider the conductivity to be almost infinite, and at high frequencies the metal can be considered almost transparent. Using it and differentiating, we have

$$\frac{F}{L^2} = -\frac{3\hbar c \pi^2}{4d^4}\sum_l \int_0^\infty f(u,l)\sqrt{u+l^2}du.$$

By selecting different functions, you can try to get the final result. For example, you can take $f(u,l) = 1/[1+l^2(u+l^2)^2]$. Then $F/L^2 = -3\hbar c \pi^3 \varsigma(3/2)/(8d^4)$, and for $f(u,l) = 1/[1+l^4(u+l^2)^2]$ we get $F/L^2 = -3\hbar c \pi^3 \varsigma(3)/(8d^4)$. The final result of Casimir is the asymptotic result of separating the finite part from infinity based on the asymptotic Euler–Maclaurin formula, and to obtain it, we had to introduce a clipping factor, which is not worth doing for infinite conductivity. The fact that the Casimir formula is obtained for extremely large $d$ is noted in [4,5]. For large $d$, it also follows from (34). For a metal, large $d$ means where plasma wavelengths are on the order of hundreds of nanometers.

Note that it is impossible to strictly go to the limit $d \to \infty$ within the framework of the used model (1) based on the approach [1, 40] for the resonator. The reason for this limitation of the model is related to the fact that we considered $d \ll D$ and then moved to the limit $D \to \infty$.

After that, you can no longer go over the limit $d \to \infty$. This corresponds to the choice $d = D/4$ and limit transition $D \to \infty$ (Fig. 1). In the case of an electric wall, the arrangement of graphene in this plane does not change the energy with an infinitesimal displacement $\Delta d$. Therefore, the contribution of modes is zero at $d \to \infty$. This is not obvious for a magnetic wall. Taking it for her $d = D/2 + \Delta d$, we get

$$\tilde{k}_z = \frac{4 \arctan \beta(\tilde{k}_z)}{D} + \frac{4l\pi}{D},$$

$$\beta(\tilde{k}_z) = -\frac{k_0 \delta y_{mn}^{e,h}}{\tan(\tilde{k}_z \Delta d / 2) + k_0 \delta y_{mn}^{e,h}}.$$

At $\Delta d = 0$ $\Delta d = 0$ we have $\beta(\tilde{k}_z) = -1$, i.e. independence from $\delta$. This means that there is no contribution to the force. Then at $D \to \infty$ we get the absence of force with infinite removal of sheets. The Van Kampen method corresponds to another $D = \infty$, and its convergence is the better the greater the $d$.

Let's note the main results of the work. At zero and small distances, the force is finite and attractive. Further, with increasing distance, it grows, reaches a maximum and decreases according to the law $1/d^4$. It is shown that the Casimir result and the Lifshitz result of force decay are valid according to the law $1/d^4$ for large distances, which is actually noted in [4]. The result [31] for plates shows a significant decrease in force in thin plates, but it also corresponds to large distances, since the method [35] with an increasing function at infinity is used. There are a number of papers where the Casimir forces for two graphene sheets are considered, for example, [20–27]. In particular, in [20, 23], the force is proportional to $1/d^4$ for any values of $d$, i.e., as in the Casimir case. There are reviews on Casimir's strength [29,41–46]. In particular, [46] provides a quantum field analysis in the Matsubara formalism. In the classical approach, the Lifshitz formula is usually used for the vacuum gap in the dielectric space [20–22,27]. This formula includes the reflection coefficients of E-modes and H-modes from a dielectric half-space of infinite thickness, and for graphene they are replaced by reflection coefficients from its sheets. An infinite half-space is not completely transparent, i.e. any wave in it, even with infinitely weak dissipation, is attenuated. The final plate is partially transparent, so in addition to the reflection coefficients, there are also transmission coefficients that are not involved. Graphene is well transparent. Although the lack of conductivity in this graphene model leads to zero reflection coefficients and zero strength, these formulas for graphene, considered as a sheet with zero thickness, still require justification. In Table 1, we compare our results with those of [20] and [24]. The results of [20] are given for the electron concentration $n = 10^{16}$ 1/m2. By virtue of the formula $n = (2/\pi)\mu_c^2 /(\hbar v_F)^2$, this corresponds to the chemical potential $\mu_c = 0.0783$ eV, which was used in the calculations. The results from [20] and [24] tend to infinity at $d<1$ nm and are clearly incorrect.

Table 1. Comparison of force density (N/m$^2$) with the results of [20] and [24]

| Results | $d=0.5$ нм | $d=1$ нм | $d=10$ нм | $d=60$ нм | $d=1000$ нм |
|---|---|---|---|---|---|
| [20] | $4\cdot 10^7$ | $2.5\cdot 10^6$ | 255 | 0.19 | $2.46\cdot 10^{-6}$ |
| [24] | $1.3\cdot 10^8$ | $8.4\cdot 10^6$ | 840.0 | 0.65 | $8.42\cdot 10^{-6}$ |
| Fig. 3 | $6.49\cdot 10^{-6}$ | $8.35\cdot 10^{-5}$ | $1.91\cdot 10^{-3}$ | $4.31\cdot 10^{-3}$ | $2.46\cdot 10^{-6}$ |
| Result 1, Fig. 5 | $5.5\cdot 10^5$ | $4\cdot 10^4$ | 125 | 0.3 | $2.71\cdot 10^{-6}$ |

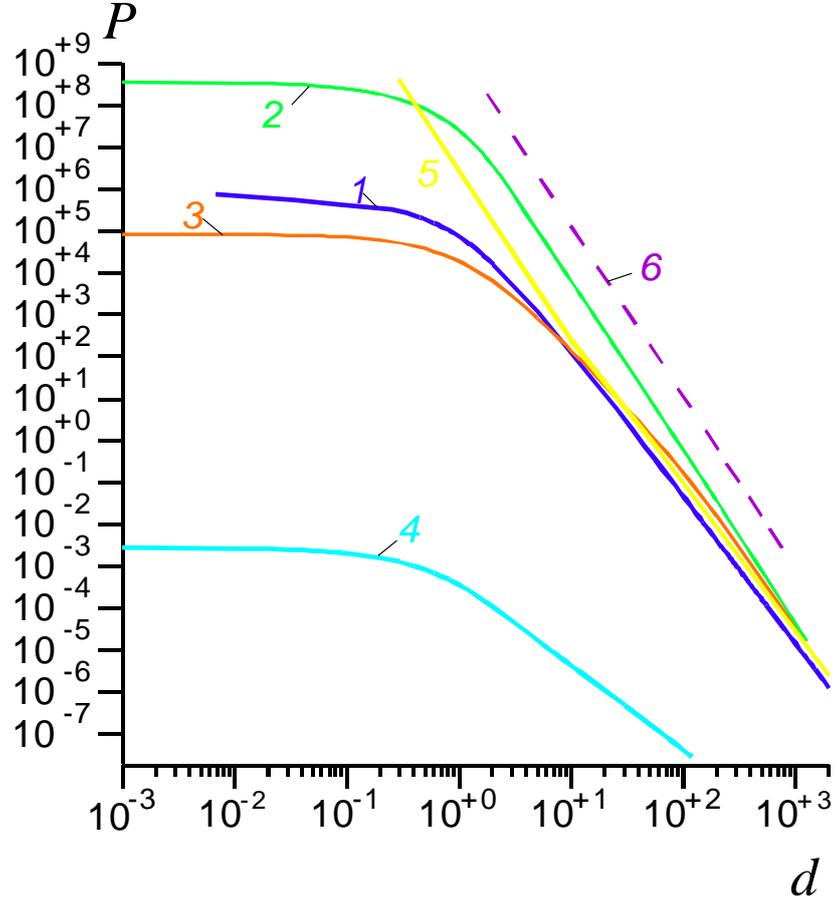

Fig. 5. The Casimir force density $F/L^2$ (N/m$^2$) between two graphene sheets as a function of the distance $d$ (nm) according to model (34) with $\mu_c = 0.0783$ (curve 1), according to model (36) without a correction factor ($\mu_c = 0.0783$ eV, curve 2) and with correction factors and $\mu_c = 0.1$ (curve 3). Line 4 is the result from [20]. Dashed line 5 is the result of [1]. $\omega_{c0} = 10^{12}$ Hz is everywhere

Figure 5 shows the results for integral (36) without a multiplier $Q_{e,h}$ and with such multiplier, as well as for the same integral (34) with a function $\tilde{\tilde{f}}_{e,h}(\kappa, k, d)$. Since the latter is equal to $\exp(2Kd)-1$ and corresponds to the Casimir problem for infinite conductivity, the results for it determine the force at large distances. In our case, it is more than $\lambda_p$. In the domain of $d > \lambda_p$, the results almost coincided with [20] (i.e., with the results of Lifshitz theory) both for curve 1, fig. 2, and for curves 1, 2, 3, as well as with the Van Kampen formula Fig. 5. At $d > 10$ nm, the matching of the correction factor model with the Van Kampen method and with Lifshitz theory is quite good. A model without a correction factor at short distances gives overestimated results. Summing the contributions of the resultant and evanescent modes practically gives the result according to the Van Kampen formula (curve 1). The method is applicable for the final temperature, for which a temperature-dependent length $\delta(\omega, T)$ should be used. In Fig. 2, dashed curve 2 and curve 3 correspond to room temperature. The temperature effect here is only due to changes in boundary conditions. However, at high temperatures, the contribution from thermal fluctuations [20] associated with the presence of a thermal field should be added.

The method is applicable to any two-dimensional conductive structure. For graphene conductivity, strict formulas should be used [14] with integration over the entire Brillouin zone. Unfortunately, such integration is possible only numerically. Formulas [14] also take into account spatial dispersion, but it is taken into account for large *k*. Spatial dispersion can lead to an additional limitation of the integrals. An energy gap may occur in graphene (for example, due to finite dimensions or doping), which must be taken into account. The results obtained give finite force densities at all distances, decreasing over long distances. They can be interpreted as follows. At infinitesimal distances, the finite conductivity of graphene does not change the conditions on the electric walls, therefore, at these distances, modes with an electric wall almost do not contribute to the force. The attractive contribution of magnetic modes with a magnetic wall at such distances can be explained by the fact that common-mode plane currents with long-wave fluctuations are almost constant, their magnetic field does not depend on distance, and such surface current densities, according to the Lorentz force, are also attracted with a force density independent of distance. An increase in *d* leads to a violation of common–mode behavior, while magnetic modes make a repulsive contribution, and electric modes make an attractive contribution. The total contribution is weakly dependent on *d*, and then begins to increase. At large distances, fluctuations become strongly local and uncorrelated, which leads to decreasing dipole-dipole attraction. Although we have obtained a separation of the contributions of radiated and evanescent modes, the Van Kampen formulas are the most convenient for calculations. They take into account both radiated modes and evanescent modes in a single way.

The results of the work are applicable to graphene sheets in a transparent, non-dissipative medium without dispersion (at the frequencies under consideration) with a dielectric constant $\varepsilon$. In this case, all resonator frequencies are formally reduced by a factor $\sqrt{\varepsilon}$. Graphene sheets in such an ideal environment can be fixed at a distance of $d$. Changing the charge concentration (Changing $\mu_c$) and controlling the force within large limits is possible by applying an electrostatic potential to the sheets. At small distances of less than 1 nm, the quantum capacity of graphene should be taken into account to correctly describe the charge density [47]. In addition to the Casimir force, there is an electrostatic attraction between the sheets as the capacitor plates interact with a density proportional to the charge density. This force decreases as $1/d^2$. The results are easy to modify for the case of a thin dielectric layer between graphene sheets, and for the interaction of graphene with a perfectly conductive surface. In the latter case, it is sufficient to use only the formulas for the electrical wall.

**Financing the work**
The work was carried out with the financial support of the Ministry of Education and Science of the Russian Federation within the framework of the state assignment (FSRR-2023-0008).

______________

**Figures**

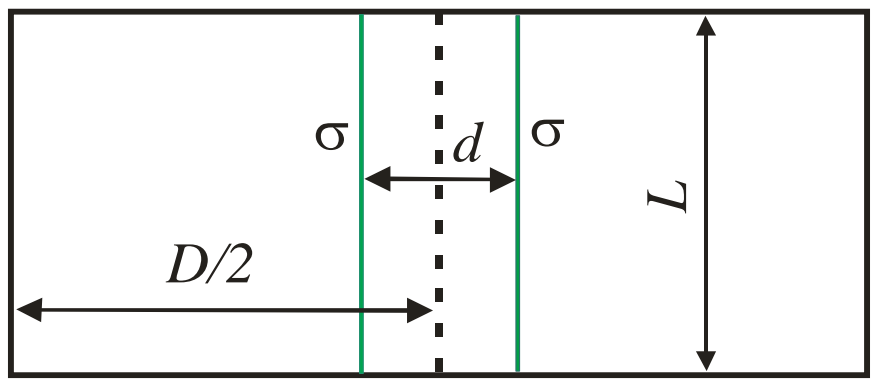

Fig. 1. Rectangular resonator with two conductive graphene sheets. The dash marks the walls: electric or magnetic

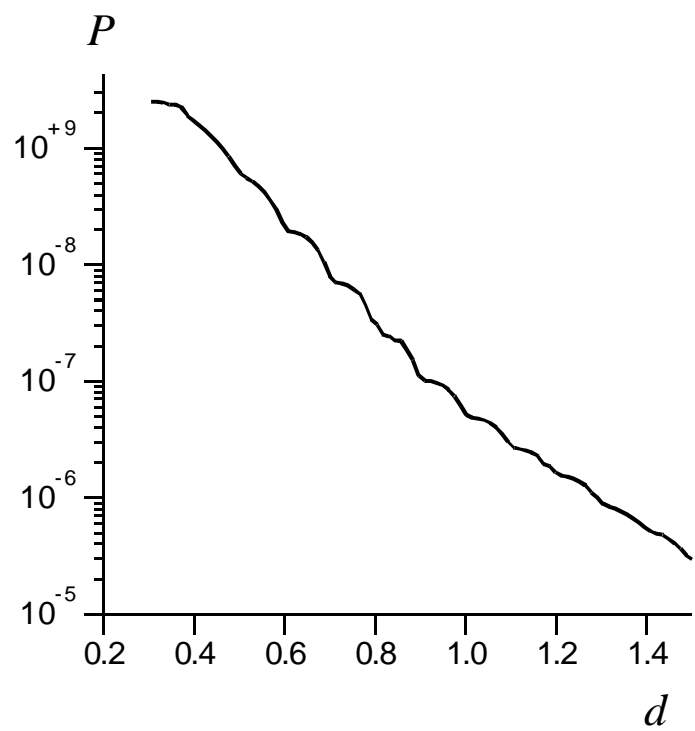

Fig. 2. Van der Waals force density $P$ (N/m$^2$) as a function of distance (nm)

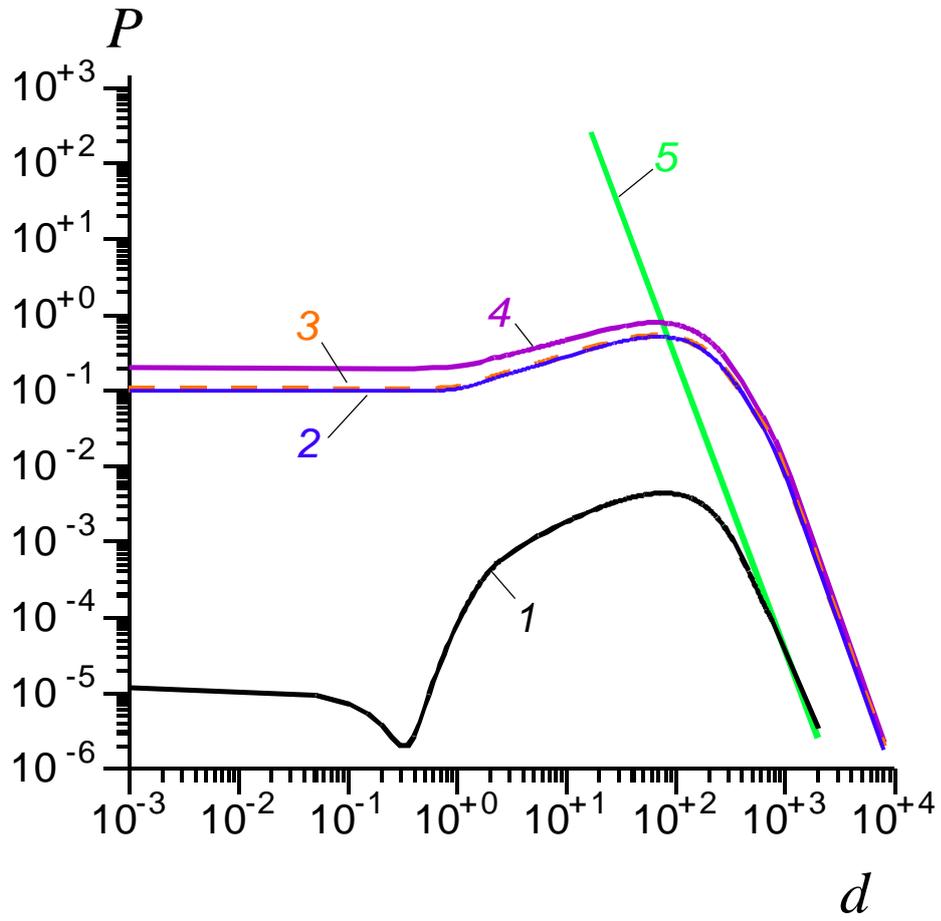

Fig. 3. The density of the Casimir attractive force $P=F/L^2$ (N/m$^2$) between two graphene sheets according to formula (14), depending on the distance $d$ (nm) at $\mu_c = 0.0783$ eV (curve 1) and $\mu_c = 7.8$ eV (curves 2–4) and different temperatures: $T=0$, (curve 2), $T=300$ K (3), $T=900$ K (4). Line 5 is the result from [20] at $T=0$. The curves are plotted for $\nu = 4$, $\omega_c = 10^{12}$ Hz

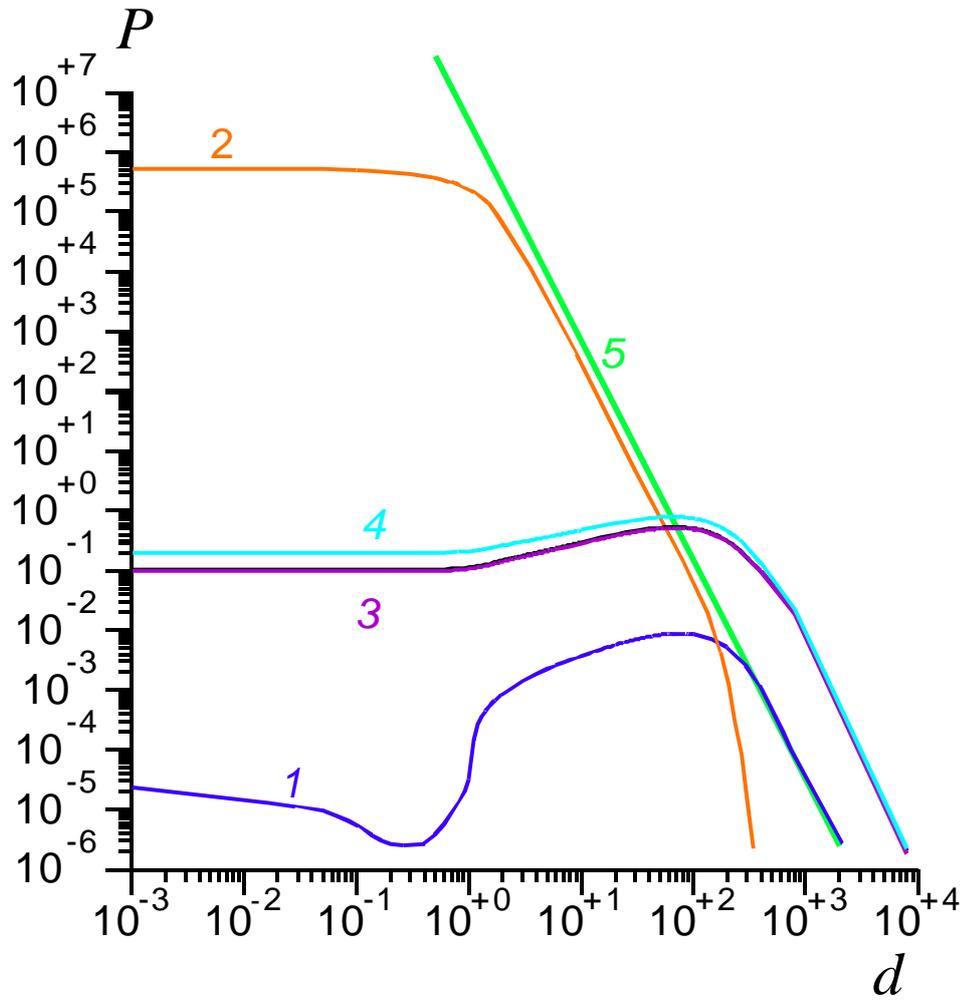

Fig. 4. Densities (N/m$^2$) of the attractive force $\tilde{P}$ (curves 1,3,4) and $P_{ev}$ (2) between two graphene sheets as a function of the distance $d$ (nm) at $\mu_c = 0.0783$ eV (curves 1,2) and $\mu_c = 7.8$ eV (curves 3,4). Curves 1–3 are plotted at a temperature of $T=600$, curve 4 is at $T=900$ (K). Line 5 is the result from [20] at $T=0$. Everywhere $\nu = 4$, $\omega_c = 10^{12}$ Hz

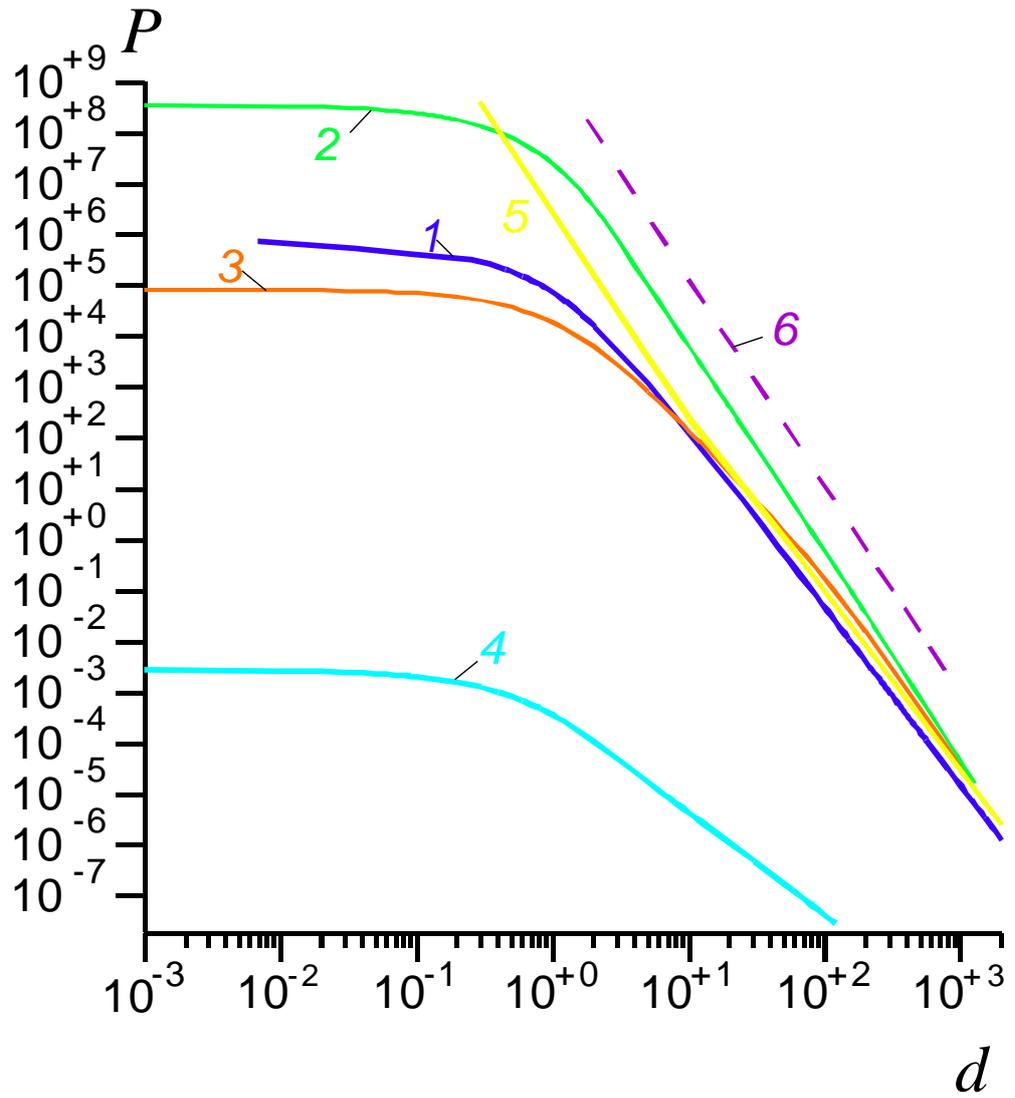

Fig. 5. The Casimir force density $F/L^2$ (N/m$^2$) between two graphene sheets as a function of the distance $d$ (nm) according to model (34) with $\mu_c = 0.0783$ (curve 1), according to model (36) without a correction factor ($\mu_c = 0.0783$ eV, curve 2) and with correction factors and $\mu_c = 0.1$ (curve 3). Line 4 is the result from [20]. Dashed line 5 is the result of [1]. $\omega_{c0} = 10^{12}$ Hz is everywhere

37